\newif\ifCHANGES
\CHANGEStrue

\documentclass[
    3p,
    sort&compress,
    times,
]{elsarticle}

\usepackage[utf8]{inputenc}

\usepackage{xparse}

\usepackage{microtype}

\usepackage{amsmath}
\usepackage{amssymb}
\usepackage{bm}
\usepackage{mathtools}
\usepackage{physics}
\usepackage{siunitx}

\usepackage{graphicx}
\usepackage{xcolor}

\usepackage[labelformat=simple]{subcaption}

\usepackage{todonotes}

\usepackage{acronym}
\usepackage{hyperref}

\definecolor{edits}{RGB}{220,0,0}
\definecolor{strike}{RGB}{150,50,50}
\ifCHANGES
    \usepackage[normalem]{ulem}
    
    \NewDocumentCommand\STRIKE{+m}{{\color{strike}\sout{#1}}}
\else
    
    \NewDocumentCommand\STRIKE{+m}{}
\fi

\NewDocumentCommand\eg{}{e.\,g.}
\NewDocumentCommand\ie{}{i.\,e.}
\NewDocumentCommand\cf{}{cf.}

\NewDocumentCommand\eqperiod{}{\,\text{.}}
\NewDocumentCommand\eqcomma{}{\,\text{,}}

\NewDocumentCommand\T{}{\mathsf{T}}
\RenewDocumentCommand\vec{m}{\bm{#1}}

\NewDocumentCommand\mat{m}{\bm{#1}}
\NewDocumentCommand\snabla{m}{\vnabla_{\!#1}}

\DeclarePairedDelimiterX\set[1]{\{}{\}}
    {#1}

\NewDocumentCommand\Ws{}{\ensuremath{\mathcal{W}_\mathrm{s}}}
\NewDocumentCommand\Wu{}{\ensuremath{\mathcal{W}_\mathrm{u}}}
\NewDocumentCommand\gammaS{}{\ensuremath{\vec{\tilde{\Gamma}}^\mathrm{s}}}
\NewDocumentCommand\gammaU{}{\ensuremath{\vec{\tilde{\Gamma}}^\mathrm{u}}}
\NewDocumentCommand\thetaS{}{\ensuremath{\theta^\mathrm{s}}}
\NewDocumentCommand\thetaU{}{\ensuremath{\theta^\mathrm{u}}}
\NewDocumentCommand\vThetaS{}{\ensuremath{v_\theta^\mathrm{s}}}
\NewDocumentCommand\vThetaU{}{\ensuremath{v_\theta^\mathrm{u}}}
\NewDocumentCommand\thetaDS{}{\ensuremath{\theta^\mathrm{DS}}}
\NewDocumentCommand\aDS{}{\ensuremath{a^\mathrm{DS}}}
\NewDocumentCommand\aDSdot{}{\ensuremath{\dot{a}^\mathrm{DS}}}
\NewDocumentCommand\RR{m}{(#1)}
\NewDocumentCommand\kAvg{}{\ensuremath{k_\mathrm{avg}}}

\NewDocumentCommand\Hk{}{\ensuremath{H_K}}
\NewDocumentCommand\Ms{}{\ensuremath{M_\mathrm{S}}}
\NewDocumentCommand\Heff{}{\ensuremath{\vec{H}_\mathrm{eff}}}
\NewDocumentCommand\Hext{}{\ensuremath{\vec{H}_\mathrm{ext}}}
\NewDocumentCommand\Han{}{\ensuremath{\vec{H}_\mathrm{an}}}
\NewDocumentCommand\Hd{}{\ensuremath{\vec{H}_\mathrm{d}}}

\ProvideDocumentCommand\Ac{m}{\ac{#1}}

\journal{Communications in Nonlinear Science and Numerical Simulation}

\begin{document}

\begin{frontmatter}
    \title{Transition state theory characterizes thin film macrospin dynamics
        driven by an oscillatory magnetic field: Inertial effects}

    \author[us]{Michael Maihöfer}
    \author[us]{Johannes Reiff}
    \author[us]{Jörg Main}
    \address[us]{
        Institut für Theoretische Physik I,
        Universität Stuttgart,
        70550 Stuttgart, Germany
    }

    \author[jhuc,jhucbe]{Rigoberto Hernandez\texorpdfstring{\corref{cor}}{}}
    \ead{r.hernandez@jhu.edu}
    \cortext[cor]{Corresponding author}
    \address[jhuc]{
        Department of Chemistry,
        Johns Hopkins University,
        Baltimore, Maryland 21218, USA
    }
    \address[jhucbe]{
        Departments of Chemical \& Biomolecular Engineering,
        and Materials Science and Engineering,
        Johns Hopkins University,
        Baltimore, Maryland 21218, USA
    }

    \date{\today}

    \begin{abstract}
        Understanding the magnetization switching process
        in ferromagnetic thin films
        is essential for many technological applications.
        We investigate the effects of periodic driving via magnetic fields
        on a macrospin system
        under explicit consideration of inertial dynamics.
        This is usually achieved by extending
        the Landau--Lifshitz--Gilbert equation
        with a term including the second time derivative of the magnetization.
        The dynamics of the magnetization switching can
        then be characterized by its switching rate.
        We apply methods from transition state theory for driven systems
        to resolve the rate of magnetization switching in this general case.
        In doing so, we find that magnetization exhibits resonance-like behavior
        under certain driving conditions, and
        it can be affected strongly by the system's relaxation rate.
    \end{abstract}

\end{frontmatter}


\acrodef{BCM}{binary contraction method}
\acrodef{DoF}{degree of freedom}
\acrodefplural{DoF}{degrees of freedom}
\acrodef{DS}{dividing surface}
\acrodef{LD}{Lagrangian descriptor}
\acrodef{LMA}{local manifold analysis}
\acrodef{MRAM}{magnetoresistive random access memory}
\acrodefplural{MRAM}{magnetoresistive random access memories}
\acrodef{NHIM}{normally hyperbolic invariant manifold}
\acrodef{PES}{potential energy surface}
\acrodef{TS}{transition state}
\acrodef{TST}{transition state theory}
\acrodef{LLG}{Landau--Lifshitz--Gilbert}


\section{Introduction}
\label{sec:intro}

Single-domain nanomagnets have been studied extensively,
in part because of their potential use in magnetic storage devices
such as \acp{MRAM}~\cite{Augustine2012a}.
These applications in spintronics require control over
the magnetization switching process in a ferromagnetic thin film.
The application of a step function change in a static magnetic field
is perhaps the simplest way to
flip or drive the magnetization of a magnetic device.
Stoner and Wohlfarth~\cite{Stoner1948a} characterized the corresponding
magnetic model while also allowing for uniaxial anisotropy,
and it is now well understood~\cite{Tannous2008a, Tannous2008b}.
Although such static magnetic fields can be used to drive macrospins,
they tend to be quite large and thus attention has recently shifted
towards more elaborate switching strategies to minimize energy requirements.
For example, microwave-assisted switching~\cite{
    Thirion2003a, Zhu2008a, Okamoto2012a, Taniguchi2014b, Suto2015a}
has been employed successfully by several groups,
where alternating magnetic fields
perpendicular to the easy axis are used
to excite the precession of the magnetic moment.
Such steps or pulses in the magnetic
fields facilitate magnetization switching along the easy axis~\cite{
    Barros2011a, Barros2013a, Klughertz2015a, Taniguchi2016a}.
Control of magnetization switching has also been achieved
using rotating radio-frequency fields perpendicular to
the easy axis~\cite{Rivkin2006a, Taniguchi2015a}.

In this paper, we focus on the macrospin dynamics driven by
a single alternating field along the easy axis
in the absence of any static external fields.
Magnetization switching requires
the crossing of the energy barrier separating two potential wells
corresponding to
\emph{spin up} and \emph{spin down}, respectively.
We report the consequences on macrospins
after their magnetization has been lifted near the barrier region
but the flip has not yet occurred.
This barrier gives rise to invariant manifolds
that determine the switching behavior
and are directly linked to the switching rate between the two potential wells.
Using a harmonic driving field, the dynamics near the barrier can be modified
to yield different switching rates.
The dynamics of such systems is usually described by the \ac{LLG} equation,
which has only a two-dimensional phase space.

Some of us used the \ac{LLG} equation within a \ac{TST} framework previously
to reveal the phase space dynamics near the energy barrier
and to obtain rates for magnetization switching when
the inertia is neglected~\cite{hern22b}.
Without driving, the macrospins can be described by
the invariant manifold of the \ac{TS},
and the stability of those spins on the \ac{TS}
is quantified by the decay rate of macrospins in their vicinity.
In a driven system, such as when torque is exerted
on the macrospin by an oscillatory field,
this manifold is time dependent, but can nevertheless
be characterized.
Here, we show that the associated decay rate can be computed using
the methods of nonequilibrium \ac{TST}~\cite{hern19a, hern21j},
even when inertial terms are included.
For simplicity, we consider only the case when
the torque exerted by the oscillatory field is
in the direction of the easy axis.

In chemical reaction theory and more general activated dynamics
problems~\cite{rmp90},
\ac{TST} has been established as a useful,
though often approximate, approach for calculating reaction rates.
It has been adopted to non-chemical problems, such as atomic
physics~\cite{Jaffe00}, solid state physics~\cite{Jacucci1984},
cluster formation~\cite{KomatsuzakiBerry99a, KomatsuzakiBerry02},
diffusion dynamics~\cite{toller, voter02b}, and
cosmology~\cite{Oliveira02, Jaffe02, hern22a}.
\Ac{TST} approaches have been used previously
by other groups~\cite{Bessarab2012a, Wang2007a}
to calculate the thermal stability of magnetic states.
As we noted above,
a time-dependent \ac{TST} can also provide decay rates
under the influence of an external oscillating driving field~\cite{hern22b}.
We extend these considerations here to address the
magnetic moment in a periodically oscillating magnetic field
using an extension of \ac{TST} for driven systems when considering inertia.
In particular, we found that
an external driving's amplitude and frequency can influence these properties.

In this paper, we now consider the \ac{LLG} equation
extended by an inertial term~\cite{Ciornei2011, Fahnle2011},
which has proven~\cite{LiBarra2015, Neeraj2021a}
to describe the magnetization dynamics at picosecond time scales.
From the dynamical point of view, the inclusion of inertia introduces
a four-dimensional phase space which can give rise to added complexity.
The extended model also features an additional material parameter,
the relaxation time $\tau$,
that can influence the rates and phase space dynamics
as a function of the driving parameters.
Here, we focus on the case of sufficiently large $\tau$
to ensure that the governing differential equations remain second order.
We further reveal
the effects of a harmonic driving field on the rate of magnetization switching
via a statistical approach using \ac{TST}.

We start by introducing the thin-film model system
of Fig.~\ref{fig:config_film} in Sec.~\ref{sec:methods/model}.
A brief overview of \ac{TST}
in the context of the magnetization switching problem
is provided in Sec.~\ref{sec:methods/tst},
including an overview of numerical methods implemented in this framework.
We demonstrate in Sec.~\ref{sec:results/trajectories}
that \ac{TST} can be applied to uniformly magnetized systems.
The results of our \ac{TST}-based analysis are reported
in Secs.~\ref{sec:results/nhim} and~\ref{sec:results/rate}.


\section{Methods and materials}
\label{sec:methods}

\subsection{Inertial terms in the driven thin-film model}
\label{sec:methods/model}

\begin{figure}
    \centering
     \includegraphics[width=0.8\columnwidth]{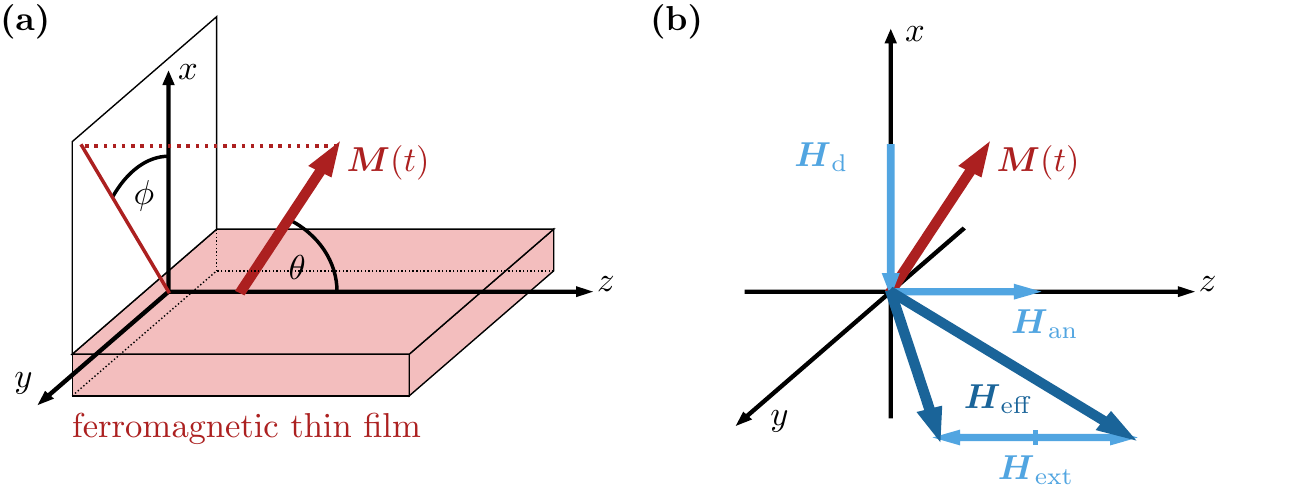}
    {\phantomsubcaption\label{fig:config_film}}
    {\phantomsubcaption\label{fig:config_fields}}
    \caption{%
        Schematic of
        \subref{fig:config_film}~a ferromagnetic thin film
        with time-dependent, uniform magnetization $\vec{M}(t)$ and
        \subref{fig:config_fields}~magnetic field components
        acting on the magnetization.
        The magnetic field \Heff\ in the ferromagnetic thin film
        consists of a demagnetizing field \Hd, an anisotropy field \Han,
        and an external driving field \Hext.
        The latter is shown as a double-arrow
        indicating its time-periodic oscillation around zero.
        The calculations are carried out in standard spherical coordinates
        with $\theta$ being the angle between $\vec{M}$ and the $z$-axis,
        and $\phi$ being the angle in the $xy$-plane relative to the $x$-axis.}
    \label{fig:config}
\end{figure}

We previously used the
Gilbert equation~\cite{Gilbert2004, Apalkov2005, Abert2019a}
to describe magnetization
while ignoring the inertial term~\cite{hern22b}.
This is justified as long as the macrospin dynamics are not too fast.
However, inertial effects play an important role for fast spin dynamics
on a time scale of the order or below picoseconds~\cite{Neeraj2021a}.
While the Gilbert equation of motion accounts for
precession (in the first term) and damping (in the second term),
inclusion of inertial effects~\cite{Ciornei2011, Fahnle2011}
manifests itself as
a third term
in the extended equation 
\begin{equation}
    \label{eq:implicit_llg}
    \dot{\vec{m}} =
        - \gamma \vec{m} \cross \Heff
        + \alpha \vec{m} \cross \dot{\vec{m}}
        + \alpha \tau \vec{m} \cross \ddot{\vec{m}}
    \eqcomma
\end{equation}
where the unit vector $\vec{m}(t)$ is the evolution of
the direction of magnetization whose magnitude is \Ms,
the parameter $\gamma$ is the gyromagnetic factor,
$\alpha$ is a non-dimensional damping parameter,
$\tau$ is the relaxation time,
and $\Heff = \Hd + \Han + \Hext$ (\cf\ Fig.~\ref{fig:config}).
In turn,
the effective magnetic field can be written as the gradient
of the free-energy potential $U$,
\ie, $\Heff = -(1 / \Ms) \snabla{\vec{m}} U$.
Curiously Eq.~\eqref{eq:implicit_llg}
arises not just from the framework of the breathing Fermi surface
model~\cite{Steiauf2005a, Faehnle2006a, Fahnle2011},
but also from a complementary
non-equilibrium thermodynamics approach~\cite{Ciornei2011}.
It has also been
benchmarked through a ferromagnetic resonance experiment~\cite{LiBarra2015}.

\begin{figure}
    \centering
    {
         \includegraphics[width=0.54\columnwidth]{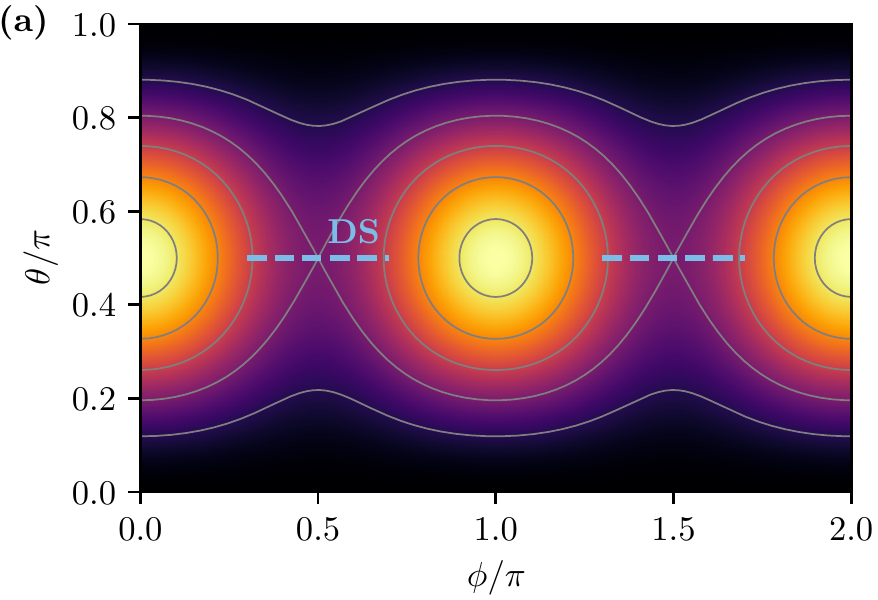}
        \phantomsubcaption\label{fig:potential}
    }
    \hfill
    {
         \includegraphics[width=0.44\columnwidth]{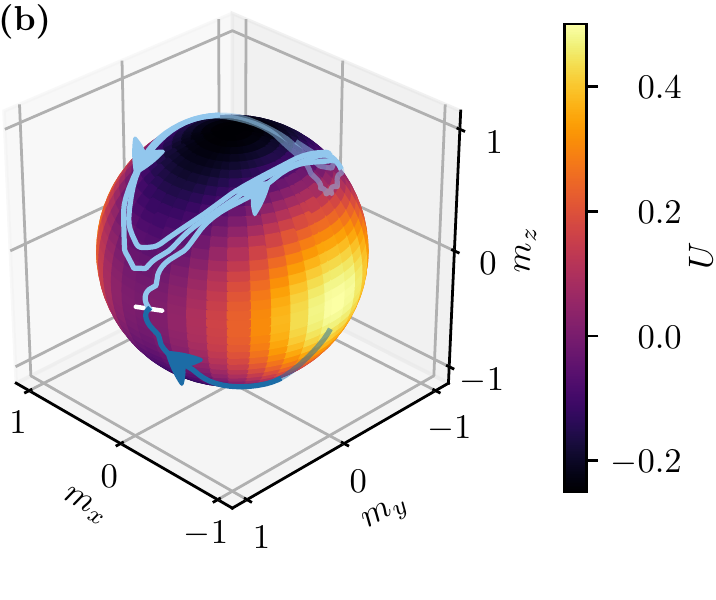}
        \phantomsubcaption\label{fig:trajectory}
    }
    \caption{%
        \subref{fig:potential}~Free-energy potential $U(\vec{m})$
        from Eq.~\eqref{eq:potential}
        at $t = 0$.
        In this representation,
        the spin-down state is at the bottom
        and the spin-up state is at the top,
        and separated by a naive \acf{DS}---shown as a dashed line---%
        located horizontally at $\theta = \pi / 2$.
        The potential exhibits two rank-1 saddle points
        at $(\phi, \theta) = (\pi / 2, \pi / 2)$ and $(3 \pi / 2, \pi / 2)$.
        \subref{fig:trajectory}~Example trajectory
        for $\tau = \num{20}$ and $H = \num{0}$.
        For the reference system in Eq.~\eqref{eq:reference_parameters},
        the potential difference between the saddle point and the minimum
        corresponds to $\Delta U = \SI{3.14e5}{\J\per\cubic\m}$ and
        relaxation time $\tau = \SI{90.21}{\ps}$.
        The trajectory is initialized at
        $\theta = \num{0.5} \pi$, $\phi = \num{0.55} \pi$,
        and $v_\theta = v_\phi = \num{-0.049}$ (white line segment)
        near the saddle point that separates spin-up and spin-down states.
        The trajectory is propagated in both forward and backward time direction.
        Arrow heads on the trajectory indicate the direction of time.
        The trajectory is drawn
        with a lighter shade in forward direction (spin-up state) and
        with a darker shade in backward direction (spin-down state).
        Transitions between the two spin states are defined by the \acs{DS},
        which in this example is crossed near the initial point.
        For reference, a sphere of radius $0.98$
        indicates the free-energy potential $U(\vec{m})$.
        Semitransparency is used to indicate
        when the trajectory is on the back side of the potential sphere.}
\end{figure}

Here, we use the free energy~\cite{Bauer2000a, Sun2000a}
of a ferromagnetic single-domain layer
with a time-dependent but uniform magnetization.
In the following, such a layer will be referred to as a \emph{thin film}
with uniform magnetization $\vec{M}(t)$, \cf\ Fig.~\ref{fig:config}.
This model includes the uniaxial anisotropy field \Han,
the demagnetization field \Hd\ inducing a shape anisotropy,
and the external driving \Hext\ of an infinite thin film.
The free-energy potential can be written as
\begin{equation}
    \label{eq:potential}
    U(\vec{m}, t) =
        - \frac{1}{2} \Hk \Ms m_z^2
        + \frac{1}{2} \Ms^2 m_x^2
        - \Ms H \sin(\omega t) m_z
    \eqcomma
\end{equation}
where the easy axis of the magnetic material lies in $z$-direction,
and the thin film lies in the $yz$-plane.
Here, $\Hk = \abs{\Han} / m_z$ is the anisotropy field,
and $H$ and $\omega$ are the amplitude and frequency
of the external driving field \Hext, respectively.
While we recognize that the thin film is small enough
to accommodate macrospins, we approximate the
shape anisotropy using the form for an infinitely
extended thin film for simplicity.
This approximation amounts to ignoring the relative
contribution of $m_z^2$ from shape anisotropy,
which in the current case is small
compared to the leading term in the potential, or which in any
case amounts to a rescaling of the coefficients of the
squared magnetization component terms.
Hence the potential includes only the additional term in
$m_x^2$ to account for shape anisotropy.
Figure~\ref{fig:potential} shows the potential from Eq.~\eqref{eq:potential}
in spherical coordinates at $t = 0$,
where the spin up state corresponds to the region around $\theta = 0$
and the spin down state to $\theta = \pi$.
For $t = 0$ there is a rank-1 saddle point at $\theta = \phi = \pi / 2$,
which oscillates periodically around this position over time.
Figure~\ref{fig:trajectory} shows an example trajectory obtained
from the equation of motion~\eqref{eq:implicit_llg} and the potential from
Eq.~\eqref{eq:potential}.

Equation~\eqref{eq:implicit_llg} can be written explicitly as
\begin{equation}
    \label{eq:extended_llg}
    \tau \ddot{\vec{m}} =
        - \frac{1}{\alpha} \vec{m} \cross \dot{\vec{m}}
        - \frac{\gamma}{\alpha} \vec{m} \cross (\vec{m} \cross \Heff)
        - \dot{\vec{m}}
        - \tau \dot{\vec{m}}^2 \vec{m}
\end{equation}
for $\tau > 0$,
and as the \ac{LLG} equation
\begin{equation}
    \label{eq:standard_llg}
    \dot{\vec{m}} =
        - \frac{\gamma}{1 + \alpha^2} \vec{m} \cross \Heff
        - \frac{\gamma \alpha}{1 + \alpha^2} \vec{m} \cross (\vec{m} \cross \Heff)
\end{equation}
for $\tau = 0$.
The important difference between
Eqs.~\eqref{eq:extended_llg} and~\eqref{eq:standard_llg}
is that the inertial term completely changes
the phase space structure of the macrospin dynamics.
Without inertial effects, the \ac{LLG} equation~\eqref{eq:standard_llg}
is a first-order differential equation for the dynamics of the
magnetic moment on a sphere, \ie, there are no independent velocities
or momenta.
In such a case, we have a two-dimensional phase space described by the
angles $(\theta, \phi)$ for the orientation of the macrospin.
When applying \ac{TST} to the magnetization switching process,
$\theta$ is the \emph{reaction coordinate} and $\phi$ formally plays
the role of a momentum coordinate~\cite{hern22b}.
However, with inertial effects, the macrospin velocity in the
second-order differential equation~\eqref{eq:extended_llg} must be taken
into account.
It extends the phase space from two
to four dimensions described by angles and angular velocities
$(\theta, \phi, v_\theta, v_\phi)$.
As a consequence, the limit of small $\tau\lesssim 1$
in Eq.~\eqref{eq:implicit_llg}
becomes non-trivial due to the discontinuous change
of the phase-space dimension.
Thus formally and numerically,
the two-dimensional phase space is appropriate for $\tau = 0$%
---\ie, without inertial effects---
as addressed in Ref.~\cite{hern22b},
and the four-dimensional phase space is valid
for sufficiently strong inertial effects when $\tau \gtrsim 1$.

For the numerical calculations, we use units where $\gamma = 1$ and $\Ms = 1$.
This corresponds to $(\gamma \Ms)^{-1}$ and \Ms\
as the units of time $t$ and magnetic field $H$, respectively.
In these units, we set parameters
\begin{equation}
    \alpha = \num{0.01}
    \qc
    \tau = \num{20}
    \eqcomma\qand
    \Hk = \num{0.5}
\end{equation}
unless stated otherwise.
For example, in the cases reported in
Refs.~\cite{Taniguchi2014b, Taniguchi2015a, Taniguchi2016a},
$\gamma = \SI{2.217e5}{\m\per\A\per\s}$
and $\Ms = \SI{1e6}{\A\per\m}$.
Our parameters then take the values
\begin{equation}
    \label{eq:reference_parameters}
    \tau = \SI{90.21}{\ps}
    \qand
    \Hk = \SI{5e5}{\ampere\per\meter}
\end{equation}
in physical units while $\alpha$ stays dimensionless.
The inclusion of inertia allows for a formal analogy to a classical spinning top,
as pointed out in Refs.~\cite{Wegrowe2012a, Kikuchi2015a}.
In Lagrangian formulation,
Eq.~\eqref{eq:extended_llg} can be written via the Lagrangian
\begin{equation}
    \mathcal{L} =
        \frac{1}{2} \qty(
            \alpha \tau \qty[\dot{\phi}^2 \sin[2](\theta) + \dot{\theta}^2]
            + \dot{\phi}^2 \cos[2](\theta))
        - U(\theta, \phi)
    \eqcomma
\end{equation}
which includes kinetic and potential energy terms,
and the Rayleigh dissipation function
\begin{equation}
    \mathcal{R} =
        \frac{\alpha}{2} \qty[\dot{\phi}^2 \sin[2](\theta) + \dot{\theta}^2]
    \eqperiod
\end{equation}
The equations of motion in Euler coordinates can be recovered
using the Euler-Lagrange equation
\begin{equation}
    \dv{t} \pdv{\mathcal{L}}{\dot{q}_i}
        - \pdv{\mathcal{L}}{q_i}
        + \pdv{\mathcal{R}}{\dot{q}_i}
    = 0
    \eqperiod
\end{equation}
In terms of the spinning top,
setting $\alpha = \tau = 0$ results in pure \emph{precession},
where trajectories follow equipotential lines of the potential $U(\theta, \phi)$.
Inclusion of damping---\ie, $\alpha \neq 0$ and $\tau = 0$---%
leads to a damped precession converging towards the minimum of the potential.
The further inclusion of inertia---\ie, $\alpha \neq 0$ and $\tau \neq 0$---%
leads to additional \emph{nutation} and
the complex dynamics addressed in this paper.


\subsection{Transition state theory}
\label{sec:methods/tst}

\subsubsection{Adapting TST to macrospin decays}
\label{sec:methods/tst/general}

The model described in Sec.~\ref{sec:methods/model}
features two metastable configurations along
the $z$-direction.
They are separated by a potential barrier oscillating around the $xy$-plane.
Any flip between the magnetization states
has to pass over the time-dependent energy barrier.
A magnetization configuration is trapped or might flip
depending on whether or not
its energy is sufficient to pass through the rank-1 saddle point.
The flipping rate is therefore highly dependent upon
the dynamics of the phase space in the vicinity of the rank-1 saddle.
The question of reaction rates between such metastable configurations
arises frequently in chemical reactions,
where \ac{TST} has been established to calculate these reaction
rates~\cite{truh79, pech81, pollak85, truh96, mill98}.
Recent work~\cite{hern10a, hern17a, hern19a, hern19e, hern20m, hern20o} has
addressed the use of \ac{TST} for driven chemical reactions which
is applied here to address the driven magnetization switching problem.

\begin{figure}
    \centering
     \includegraphics[width=0.8\columnwidth]{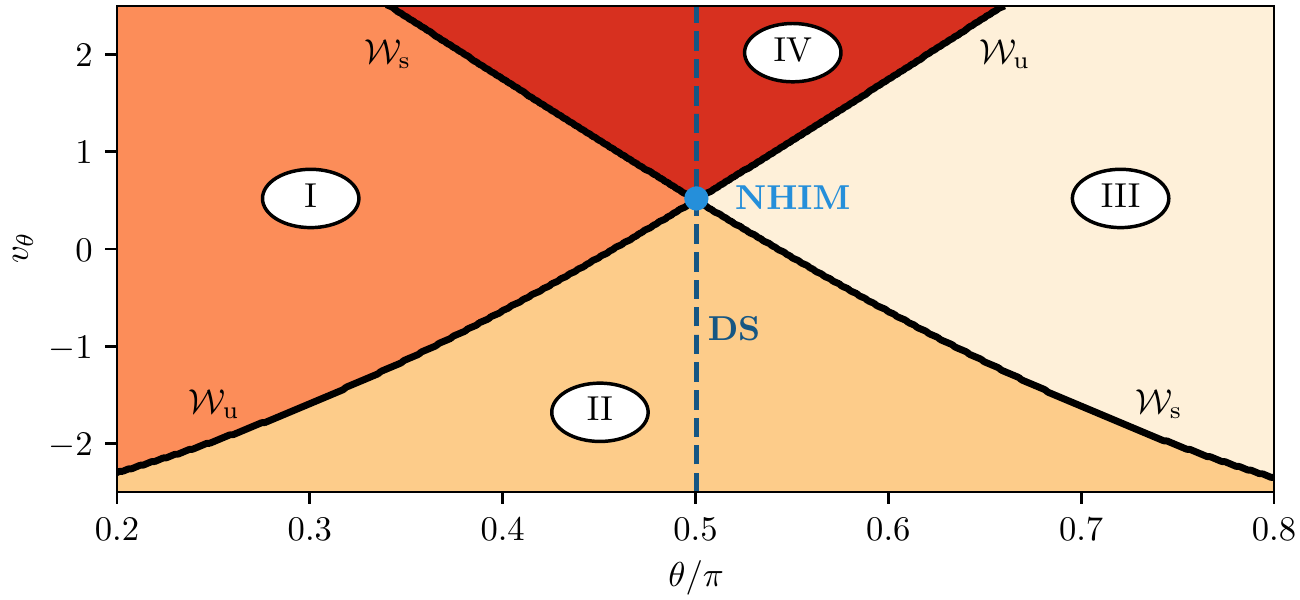}
    \caption{%
        Stable and unstable manifolds (black lines) in a phase space section
        ($\phi = \pi / 2$, $v_\phi = 0$)
        dividing the phase space into distinct regions
        \RR{I}--\RR{IV} (see text).
        The intersection of stable and unstable manifolds is the
        \acl{NHIM} (\acs{NHIM}, circle marker),
        which is the anchor for a \acs{DS} (blue dashed)
        in $v_\theta$-direction.
        Made for $\tau = \num{20}$, $H = \num{0.15}$,
        and $\omega = \pi / 8$ at $t = 0$.
        For the reference parameters in Eq.~\eqref{eq:reference_parameters}
        this corresponds to $\tau = \SI{90.21}{\ps}$, $H=\SI{1.5e5}{\A\per\m}$, and $\omega=\SI{0.547}{THz}$.}
    \label{fig:manifolds}
\end{figure}

Given a potential energy function and an equation of motion,
\ac{TST} is concerned with constructing a recrossing-free \ac{DS}
separating spin-down and spin-up basins in phase space.
The rate is then calculated by the directional flux through the \ac{DS}.
The resulting instantaneous decay rate is
\begin{equation}
    \label{eq:population_rate}
    k(t) = - \frac{\dot{N}(t)}{N(t)}
    \eqcomma
\end{equation}
where $N(t)$ is the time-dependent reactant population.
In the original formulation, the \ac{DS} was fixed, planar and orthogonal
to the reactive direction.
While a useful approximation, it overestimates the rates because of
recrossings and neglect of the stochastic forces.

\subsubsection{Normally hyperbolic invariant manifolds for macrospins}
\label{sec:methods/tst/nhim}

A major advance in improving \ac{TST}
has arisen from the identification of geometric properties of the phase space
to generalize the \ac{DS} through the association of the rank-1 saddle point
with a \acf{NHIM}~\cite{Lichtenberg82, hern93b, hern93c,
wiggins01,Uzer02,Ott2002a, wiggins16}.
This $(2 d - 2)$-dimensional manifold consists of
all trajectories trapped forever in the saddle point region.
In the following, these trajectories will be referred to as
\emph{\ac{TS} trajectories}.
It is complemented by the $(2 d - 1)$-dimensional
stable and unstable manifolds \Ws\ and \Wu\
consisting of trajectories converging to the \ac{NHIM}
for $t \to \infty$ and $t \to -\infty$, respectively.
Figure~\ref{fig:manifolds} shows a two-dimensional cut
of the magnetization's four-dimensional phase space,
reducing stable and unstable manifold to a line and the \ac{NHIM} to a point.
The stable and unstable manifolds divide the phase space
into four distinct regions, where
\RR{I}~the spin stays up,
\RR{II}~the spin flips from down to up,
\RR{III}~the spin stays down, and
\RR{IV}~the spin flips up to down.
A recrossing-free \ac{DS} must only slice through
the regions~\RR{II} and~\RR{IV},
and thus has to be anchored at the \ac{NHIM}.

The problem of constructing a recrossing-free \ac{DS}
thus reduces to finding the \ac{NHIM}.
Different numerical algorithms exist for this task,
\eg, those based on Lagrangian descriptors and variants thereof~\cite{
    Mancho2010, Mancho2013, hern15e, hern19a}.
Here, we choose the \ac{BCM}:
Any initial point can be classified numerically
using an interaction region around the saddle,
where trajectories leaving it
are clearly classified as either spin-up or spin-down configurations.
This region does not necessarily have to be defined
via the unstable direction of the saddle (\ie, $\theta$)
as long as it enables a clear classification.
In practice, we found that choosing
$\phi \in [\num{0.1} \pi,\num{0.9} \pi]$
allowed such classification.
The resulting binary classification can be performed
for the forward and backward time direction,
thus yielding the four distinct regions \RR{I}--\RR{IV} introduced above.
Since the \ac{NHIM} is of codimension 2,
the phase space can always be reduced to
an effectively two-dimensional problem
using cuts transverse to the \ac{NHIM},
as illustrated in Fig.~\ref{fig:manifolds}.

Several numerical approaches for using the \ac{NHIM} to calculate rates
are now available:
the ensemble method~\cite{hern19e, hern20m, hern21a},
the Floquet method~\cite{hern14f},
and the \ac{LMA}~\cite{hern19e, hern22a}.
Conceptually, all of these methods work by
initializing an appropriate ensemble on one side of the \ac{DS}
and counting the number of crossings over time.
The methods, however, differ significantly
in the way the ensemble is modeled mathematically
and propagated computationally.
In the cut shown in Fig.~\ref{fig:manifolds},
the \ac{NHIM} is a point of intersection
at the closure of all four regions,
and is obtained here, and throughout the present work,
using the \ac{BCM}~\cite{hern18g}.

\subsubsection{Decay rates between macrospin states}
\label{sec:methods/tst/rates}

There exist various methods for determining
the decay rate of a \ac{TS} trajectory $\vec{\Gamma}_0$.
In this paper, we adopt the \ac{LMA}
because we had found earlier that it provides a good compromise
between computational efficiency and the extent of data
that it readily provides.
The computationally expensive step is usually the
propagation of many individual states through the \ac{DS}.
For sufficiently smooth manifolds
and a linear \ac{DS} along the reactive velocity
(as satisfied in the problems addressed here),
the propagation can be replaced by an analytical calculation
revealing the stability of the system near the \ac{NHIM}.
In the following, without loss of generality,
we choose coordinates such that
the \ac{TS} trajectory $\vec{\Gamma}_0$ is always located at the origin.
Furthermore, we assume
that the \ac{DS} is parallel to the $v_\theta$-axis,
\cf\ Fig.~\ref{fig:manifolds}.

Specifically, we first obtain the Jacobian
\begin{equation}
    \mat{J} =
        \pdv
            {(\dot{\theta}, \dot{\phi}, \dot{v}_\theta, \dot{v}_\phi)}
            {(\theta, \phi, v_\theta, v_\phi)}
\end{equation}
for the four-dimensional phase space of the equations of motion
with $v_\theta = \dot{\theta}$ and $v_\phi = \dot{\phi}$.
Two of the eigenvectors of $\mat{J}$ \emph{locally} point
in the direction of the stable and unstable manifolds
while the other two are tangent to the \ac{NHIM}.
In the following calculations, we consider
a $(\theta, v_\theta)^\T$-section of phase space
with fixed $\phi$, $v_\phi$ and time $t_0$.
In this two-dimensional frame, the eigenvector pointing in the stable
direction is $\gammaS = (\thetaS, \vThetaS)^\T$, while the eigenvector pointing
in the unstable direction is $\gammaU = (\thetaU, \vThetaU)^\T$.
We now consider an equidistant, linear ensemble on the reactant side,
parameterized by
\begin{equation}
    \label{eq:ensemble}
    \vec{\tilde{\Gamma}}(a, t) = -\gammaS(t) + a \gammaU(t)
\end{equation}
with $a \in [0, 1]$.
The ensemble is initialized in a $(\theta, v_\theta)^\T$-section of phase space
with fixed bath coordinates $\phi$ and $v_\phi$,
and thus depends implicitly on these latter coordinates.
The norms of the eigenvectors $\gammaS(t_0)$ and $\gammaU(t_0)$
at initial time $t_0$ are chosen such that
$\tilde{\vec{\Gamma}}(1, 0) = \gammaS(t_0) + \gammaU(t_0)$
is located on the \ac{DS}.
The \ac{DS} can be described by a normal vector $\vec{\tilde{n}} = (1, 0)^\T$
in the section of the ensemble.
Furthermore, we define $\gammaS(t)$ and $\gammaU(t)$ as
the time evolution of states
initially located at $\gammaS(t_0)$ and $\gammaU(t_0)$, respectively.

The value of $\aDS(t)$, at which the ensemble pierces the \ac{DS},
depends on the length of $\gammaS(t)$ and $\gammaU(t)$
and is initially $\aDS(t_0) = 1$.
Since the number of reacted particles in the ensemble
is proportional to \aDS,
Eq.~\eqref{eq:population_rate} becomes
\begin{equation}
    k(t_0) = - \aDSdot(t_0)
    \eqperiod
\end{equation}
To determine $\aDSdot(t_0)$ we use the fact that
\begin{equation}
    \vec{\tilde{n}} \vdot \vec{\tilde{\Gamma}}\qty(\aDS(t), t) \equiv 0
\end{equation}
by definition.
Differentiating this identity relation with respect to time yields
\begin{equation}
    \label{eq:lma_derivation}
    \begin{split}
        0
        &= \vec{\tilde{n}} \vdot \dv{t} \vec{\tilde{\Gamma}}\qty(\aDS(t), t) \\
        &= \vec{\tilde{n}} \vdot \qty[
            - \dv{\gammaS(t)}{t}
            + \dv{\aDS(t)}{t} \gammaU(t)
            + \aDS(t) \dv{\gammaU(t)}{t}
        ]
        + \vec{\tilde{n}} \vdot \qty[
            \pdv{\vec{\tilde{\Gamma}}\qty(\aDS(t), t)}{\phi} \dv{\phi(t)}{t}
            + \pdv{\vec{\tilde{\Gamma}}\qty(\aDS(t), t)}{v_\phi} \dv{v_\phi(t)}{t}
        ]
        \eqperiod
    \end{split}
\end{equation}
Here, the $\dd{\phi(t)} / \dd{t}$ and $\dd{v_\phi(t)} / \dd{t}$ terms describe
how the ensemble moves out of the initial plane during propagation.
We can discard $\aDS(t)$
by evaluating Eq.~\eqref{eq:lma_derivation} at time $t_0$.
Using $\thetaU = \thetaS$
and the linearized equation of motion
{$\dv*{\vec{\Gamma}}{t} = \mat{J} \vec{\Gamma}$}
in the full four-dimensional phase space,
Eq.~\eqref{eq:lma_derivation} can be rewritten as
\begin{equation}
    \label{eq:lma_diff_a}
    \aDSdot(t_0) =
        - J_{\theta, v_\theta} \frac{\vThetaU - \vThetaS}{\thetaU}
        + \frac{1}{\thetaU} \dv{\thetaDS}{t}
\end{equation}
with $\dd{\thetaDS} = \vec{\tilde{n}} \vdot
    [(\partial\vec{\tilde{\Gamma}} / \partial\phi) \dd{\phi}
    + (\partial\vec{\tilde{\Gamma}} / \partial v_\phi) \dd{v_\phi}]$.
The first term on the right hand side of Eq.~\eqref{eq:lma_diff_a}
can be interpreted as
the difference in the slopes of stable and unstable manifolds in the plane.
The second term takes care of the ensemble drifting out of plane,
\ie, the change of the position of the \ac{DS}
when changing the orthogonal modes $\phi$ and $v_\phi$.

The resulting rate
\begin{equation}
    \label{eq:rate_numeric}
    k(\phi, v_\phi, t) =
        \frac{\vThetaU - \vThetaS}{\thetaU}
        - \frac{\thetaDS(t_0 + \delta t) - \thetaDS(t_0)}{\thetaU \delta t}
    \eqcomma
\end{equation}
once again, only depends on the difference of slopes $v_\theta / \theta$
of the linearized stable and unstable manifolds in plane
and the movement of the \ac{NHIM} out of plane.
For the numerical rate calculations in this paper,
we first find trajectories on the \ac{NHIM}
using the methods described in Sec.~\ref{sec:methods/tst/nhim}.
Given a specific point ($\phi$, $v_\phi$) on the \ac{NHIM} at time $t_0$,
we calculate the difference of the stable and unstable manifolds' slopes
$(\vThetaU - \vThetaS) / \thetaU$
in a two-dimensional phase-space cut transverse to the \ac{NHIM},
\ie, the reaction coordinates $\theta$ and $v_\theta$.
A trajectory is propagated for time $\delta t$
starting at the current location of the \ac{NHIM}
$(\theta^\ddagger, \phi, v_\theta^\ddagger, v_\phi)$ at time $t_0$.
Due to the definition of the \ac{DS},
the change $\theta^\ddagger(t_0 + \delta t) - \theta^\ddagger(t_0)$
corresponds to the term $\thetaDS(t_0 + \delta t) - \thetaDS(t_0)$
in Eq.~\eqref{eq:rate_numeric}.

The parameters of the \ac{LLG} equation enter the rate indirectly
by specifying the exact geometry of the \ac{NHIM}, and
the stable and unstable manifolds.


\section{Results}
\label{sec:results}

\subsection{Application of TST}
\label{sec:results/trajectories}

The inclusion of dissipation and gyroscopic effects
creates additional challenges to
the application of \ac{TST} to
magnetization switching~\cite{hern22b}.

Figure~\ref{fig:trajectory} shows a typical trajectory.
The magnetization vector moves towards
alignment with the $z$-axis.
This trajectory approximately follows equipotential lines,
with the Gilbert damping relaxing the trajectory
to the minimum of the potential.
The additional inertia creates nutation loops
that can be seen as wiggles along the general path of the trajectory.
In the forward and backward time direction,
a typical trajectory loses and gains energy, respectively, in doing so.
Thus the motion of the magnetization in the backward direction
would lead to ever growing nutation loops.

We found that the dynamics of the magnetization satisfies
the necessary assumptions for \ac{TST}
only when we consider cases in which the
relaxation times exceed a certain threshold
(here $\tau \gtrsim \num{1}$, depending on the driving parameters).
The reliable classification of trajectories into spin-up and spin-down domains
is a central requirement for
the numerical determination of the \ac{NHIM}'s position using the \ac{BCM}.
This is only possible when the dynamics is
sufficiently dominated by a bottleneck that gives rise to a
well-defined \ac{DS}.
For $\tau \lesssim \num{1}$, however,
we found that the predominant nutation circles do not admit such a structure
or a clear classification as spin-up or spin-down states.
The classification in backward time direction
is impossible in this parameter regime, for example.
Therefore, all of the structure of the reaction geometry
and the associated rate calculations reported here are
limited to cases in which the
relaxation times are above the observed relaxation threshold
($\tau \gtrsim \num{1}$),
as noted in Section 2.1.
For the purposes of this paper,
this is not a significant limitation
since recent studies~\cite{Neeraj2021a} suggest relaxation times $\tau$
on the order of \numrange{2}{20}
when specified in the dimensionless units of this work.


\subsection{Dynamics on the NHIM}
\label{sec:results/nhim}

For a system with $\tau = \num{0}$ described by
the standard \ac{LLG} equation~\eqref{eq:standard_llg},
the \ac{NHIM} is a single trajectory.
However, for $\tau > \num{0}$, described by Eq.~\eqref{eq:extended_llg},
the \ac{NHIM} becomes a two-dimensional, time-dependent manifold,
and therefore the dynamics on the \ac{NHIM}
needs to be considered in calculating the rate.

For any set of parameters,
the dynamics on the \ac{NHIM} can be visualized using Poincaré cuts.
We found that any initial starting point on the \ac{NHIM} will converge
towards a single periodic trajectory on the \ac{NHIM} over time
due to the damping terms.
Indeed, though not shown explicitly,
across the parameter regimes observed in this work,
none of the trajectories exhibited chaotic motion.
The rate at which points on the \ac{NHIM}
converge towards this periodic trajectory
scales approximately linearly with the inverse of the relaxation time $\tau$.
The resulting \ac{TS} trajectory has the
same frequency as the driving field.
The amplitude of its motion
depends on both the frequency and amplitude of the driving field.
The frequency at which
the amplitude of motion is maximal~\cite{Olive2015a, Neeraj2021a}
is approximately given by
\begin{equation}
    \label{eq:resonant_condition}
    \omega = \frac{\sqrt{1 + \alpha \tau \gamma H}}{\alpha \tau}
    \eqperiod
\end{equation}

\begin{figure}
    \centering
    \includegraphics[width=\columnwidth]{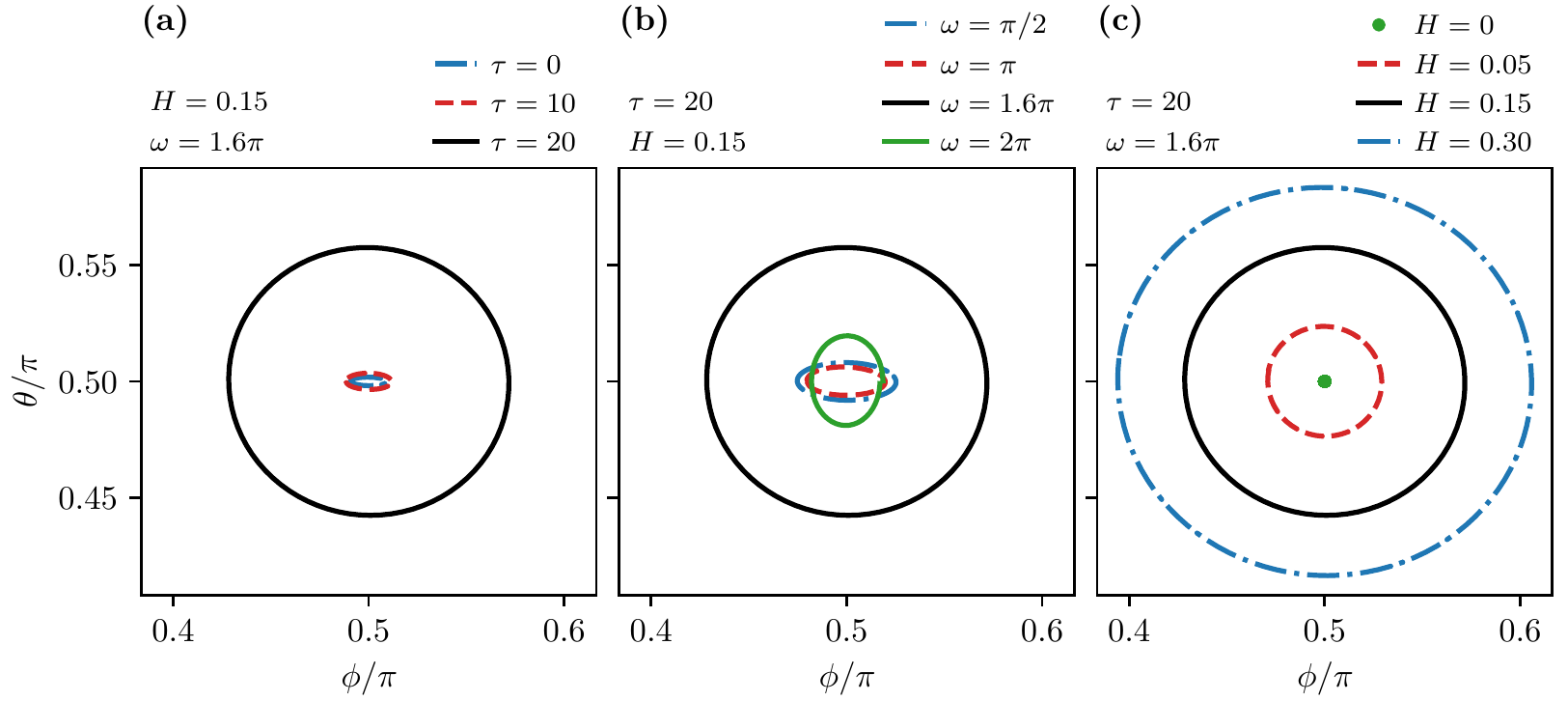}
    {\phantomsubcaption\label{fig:nhim_tau}}
    {\phantomsubcaption\label{fig:nhim_omega}}
    {\phantomsubcaption\label{fig:nhim_h}}
    \caption{%
        Periodic \acs{TS} trajectories on the \acs{NHIM}.
        We use a reference parameterization
        ($H = \num{0.15}$, $\omega = \num{1.6} \pi$, and $\tau = \num{20}$)
        shown as a black solid curve in all three panels.
        For the reference system in Eq.~\eqref{eq:reference_parameters},
        this corresponds to $H = \SI{1.5e5}{\A\per\m}$,
        $\omega = \SI{7.0}{\THz}$,
        and $\tau = \SI{90.21}{\ps}$.
        All other curves correspond to different parameterizations
        across the three panels.
        \subref{fig:nhim_tau}~At fixed
        $H = \num{0.15}$ and $\omega = \num{1.6} \pi$,
        the relaxation time $\tau$ is varied.
        The trajectory for $\tau = \num{0}$ was determined
        via the standard \acs{LLG} equation.
        \subref{fig:nhim_omega}~At fixed
        $\tau = \num{20}$ and $H = \num{0.15}$,
        the driving frequency $\omega$ is varied.
        \subref{fig:nhim_h}~At fixed
        $\tau = \num{20}$ and $\omega = \num{1.6} \pi$,
        the driving amplitude $H$ is varied.}
    \label{fig:nhim}
\end{figure}

Figure~\ref{fig:nhim} shows the periodic \ac{TS} trajectory
for different sets of parameters.
In Fig.~\ref{fig:nhim_tau}, the relaxation time is varied
at fixed driving frequency and driving amplitude.
For most values of the relaxation time, the \ac{TS} trajectory
is close to the \ac{TS} trajectory of the system without inertia.
At $\tau = \num{20}$,
the resonant condition~\eqref{eq:resonant_condition} is met and
the trajectory exhibits a particularly large amplitude.
In Fig.~\ref{fig:nhim_omega} the driving frequency is varied
at fixed driving amplitude and relaxation time.
The amplitude of the oscillation highly depends on the frequency.
In the limit of low frequency,
the \ac{TS} trajectory converges towards the trajectory of
the saddle point on the free-energy potential~\eqref{eq:potential}.
In the limit of high frequency, the \ac{TS} trajectory shrinks to a point.
In the intermediate range, the \ac{TS} trajectory exhibits resonant behavior,
increasing the amplitude of the oscillation to a maximum.
In Fig.~\ref{fig:nhim_h} the driving amplitude is varied
at fixed driving frequency and relaxation time.
The amplitude of the oscillation
increases strictly monotonically with driving amplitude.


\subsection{Rates}
\label{sec:results/rate}

We now calculate the instantaneous and average
rates for magnetization switching
associated with the \ac{TS} trajectory.
The instantaneous rates depend on time because
they are evaluated at a time-dependent position relative
to the time-dependent location of the \ac{TS}
while experiencing a time-dependent driving force;
\cf~Eq.~\eqref{eq:rate_numeric}.

We found, though not shown, that
the rate over time is periodic with the driving field
with twice the frequency of driving $\omega = 2 \pi / T$
and an average value dependent on frequency and amplitude of the driving.
This could have been expected because
the rate over time depends on the geometry of the system's phase space
along the trajectory that carries the \ac{DS}.
For the periodic \ac{TS} trajectory,
the rate is therefore periodic with the driving by construction.
Furthermore, the system exhibits the invariance
\begin{equation}
    \Heff(\theta, \phi, t) \vdot \vu{e}_j
    = -\Heff(\pi - \theta, \pi - \phi, t + T / 2) \vdot \vu{e}_j
    \eqcomma
\end{equation}
where $\vu{e}_j$ is one of the unit vectors $\vu{e}_\theta$ or $\vu{e}_\phi$.
Together with the \ac{TS} trajectory's related symmetry
(\cf\ Fig.~\ref{fig:nhim})
\begin{equation}
    \vec{m}(\theta, \phi, t) = \vec{m}(\pi - \theta, \pi - \phi, t + T / 2)
    \eqcomma
\end{equation}
this imposes further constraints on the rate,
forcing $k(t)$ to be periodic with twice the period of the driving.
Similar observations have previously been made
in other physical systems~\cite{hern19e, hern21a, hern22b}.

\begin{figure}
    \centering
    \includegraphics[width=\columnwidth]{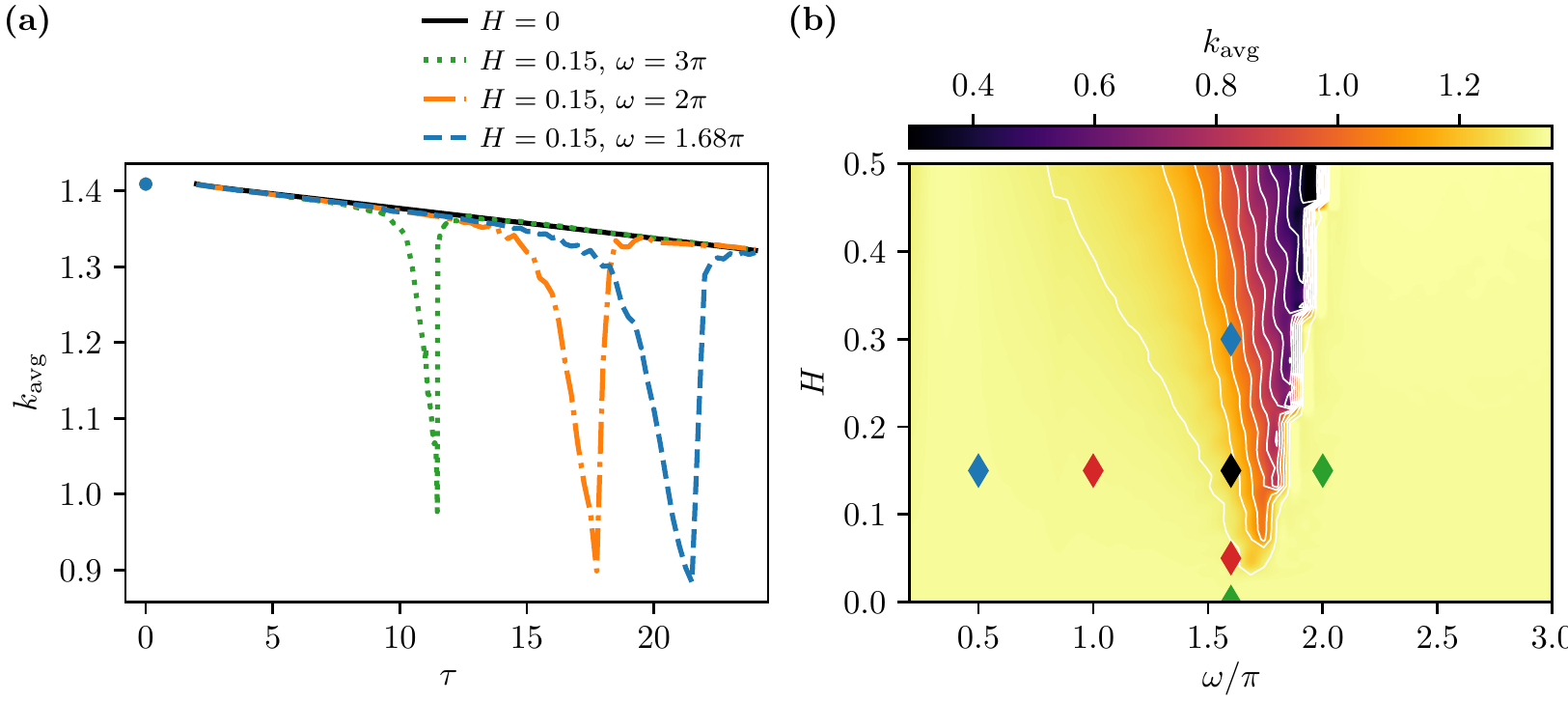}
    {\phantomsubcaption\label{fig:rates_tau}}
    {\phantomsubcaption\label{fig:rates_driving}}
    \caption{%
        \subref{fig:rates_tau}~Averaged rate \kAvg\ as a function of
        relaxation time $\tau$.
        Overlapping circle markers at $\tau = \num{0}$ show rates
        determined via the standard \acs{LLG} equation.
        \subref{fig:rates_driving}~Averaged rate \kAvg\ as a function of
        driving frequency $\omega$ and amplitude $H$ at $\tau = \num{20}$.
        The diamonds and their colors (or shades in print)
        at $H = \num{0.15}$ and $\omega = \num{1.6} \pi$
        correspond to the \acs{TS} trajectories
        in Figs.~\ref{fig:nhim_omega} and~\ref{fig:nhim_h}, respectively.
        The black diamond corresponds to $H = \SI{1.5e5}{\A\per\m}$,
        $\omega = \SI{7.0}{\THz}$,
        and $\tau = \SI{90.21}{\ps}$
        for the system in Eq.~\eqref{eq:reference_parameters}.}
    \label{fig:rates}
\end{figure}

Figure~\ref{fig:rates_driving} shows the averaged rate
as a function of driving frequency and amplitude.
The averaged rate is nearly constant for low frequencies.
Nevertheless, the rate depends significantly on the
features of the transition state trajectory in response
to the driving.
For a given $\omega$,
the average rate decreases with increasing $H$
across the parameter space shown in Fig.~\ref{fig:rates_driving}.
At driving frequencies approaching $\omega = \num{1.68} \pi$,
the rate reaches a minimum.
A slight increase in the frequency
results in a sharp increase of the rate
as the driven rate converges on the static rate.
The minimum of the potential thereby coincides with
the resonance maximum of the periodic \ac{TS} trajectory.
At the frequency of the minimum of the rate,
the transition state trajectory has a large amplitude,
and thus explores a larger phase space region
further away from the region around $\theta = \phi = \pi / 2$.
This connection is indicated by the diamonds in Fig.~\ref{fig:rates_driving},
whose location corresponds to the parameters of the external driving
used in Figs.~\ref{fig:nhim_omega} and~\ref{fig:nhim_h}.
The sharp increase in rate for incrementally larger frequency
is accompanied by the abrupt decline of the \ac{TS} trajectory's diameter.

Figure~\ref{fig:rates_tau} shows the averaged rate over relaxation time.
Without any external driving,
the averaged rate decreases almost linearly with relaxation time.
In the case of external driving, the rates stay close to
the  rate without driving for very low and high relaxation times.
However, for each specified $H$ and $\omega$,
there is a relaxation time $\tau$ where the rate is especially low
as shown in Fig.~\ref{fig:rates_tau}.
That resonance is also observed for a specific $\tau$ over the domain
of $\omega$ and $H$, such as that seen
in Fig.~\ref{fig:rates_driving} for $\tau = \num{20}$.
The minima of the rate for different $\tau$ are determined by
the resonant condition~\eqref{eq:resonant_condition}.
Finally, Fig.~\ref{fig:rates_tau} also shows that
the rates for $\tau \to \num{0}$ are consistent with
the rates calculated using the standard \ac{LLG} equation,
which is equivalent to $\tau = \num{0}$.
This is numerically nontrivial due to
the different dimensionalities of the phase space of
the equation with and without inertia,
and confirms the applicability and
accuracy of our numerical \ac{TST} implementation.


\section{Conclusion and outlook}
\label{sec:conclusion}

In this paper, we report the use of \ac{TST}
to calculate rates of magnetization switching
in a model system
considering inertial terms
and under the influence of external driving fields.
We employed a modified version of the \ac{LLG} equation
in a free-energy potential
featuring a periodically driven rank-1 saddle as our model system.
We used methods from \ac{TST} to identify the \ac{NHIM}
and to calculate the rate of magnetization flips.
We successfully identified the \ac{NHIM} for cases
with large relaxation times.

In this work, we confirmed
the use of \ac{TST} in the context of the macrospin problem
by demonstrating that it recapitulates several known properties
of driven macrospins.
While the amplitude of spin trajectories is highly dependent on
the amplitude and frequency of the driving,
we find that the system
did not exhibit chaotic behavior in the regimes considered here.
Moreover, the \ac{NHIM} contains a periodic orbit
that allows us to characterize properties of the macrospin dynamics.
For example, this periodic \ac{TS} trajectory exhibits a resonance---%
\ie, a non-monotonic dependence
of the width of the \ac{TS} trajectory on the frequency of the driving---%
which might have been expected from prior work in the
literature (cited above).
Our results are also consistent with
those obtained using the standard \ac{LLG} equation
in the limit of vanishing relaxation time.
While this limiting result may be expected from a physical point of view,
it is still an important indicator that the numerical implementation is correct
because the equations with and without inertia
correspond to phase spaces
with a different number of dimensions.

We found that any initial point on the \ac{NHIM}
converges towards the periodic \ac{TS} trajectory.
This confirms that the \ac{TST} structure is useful
in identifying stable regions in the context of the decay of macrospins.
While we did not find bifurcations in the rate over the
parameter space explore here,
we can conjecture their existence
at more extreme conditions given our earlier findings
in the context of chemical reactions~\cite{hern20m, hern21h}.

We also calculated the
instantaneous rates along the periodic \ac{TS} trajectory.
These rates have twice the frequency as the external driving.
Their values drop significantly at
specific combinations in frequency and inertia
when the periodic \ac{TS} trajectory is in resonance.
The inclusion of inertia therefore allows for richer dynamics on the \ac{NHIM}
due to increased dimensionality of the phase space with the inclusion
of the momentum variables.
Thus the average rate of magnetization switching
exhibits complex responses to external driving
in correspondence with the changing phase space structure
of the \ac{TS} trajectory
that is only available through the use of
the nonequilibrium \ac{TST} formalism.
The verification of this phenomenon is also
a challenge that remains to be done.
For practical applications,
the inclusion of spin-transfer torque is relevant,
and the \ac{TST} formalism can be extended to include such effects.
The investigation of such effects remains to be done.


\section*{Declaration of competing interest}

The authors declare that they have
no known competing financial interests or personal relationships
that could have appeared to influence the work reported in this paper.


\section*{CRediT authorship contribution statement}

\textbf{Michael Maihöfer:}
    Methodology,
    Software,
    Formal analysis,
    Investigation,
    Writing -- Original Draft,
    Visualization.
\textbf{Johannes Reiff:}
    Methodology,
    Software,
    Validation,
    Resources,
    Data Curation,
    Writing -- Review \& Editing,
    Visualization.
\textbf{Jörg Main:}
    Conceptualization,
    Methodology,
    Formal analysis,
    Resources,
    Writing -- Review \& Editing,
    Supervision,
    Project administration,
    Funding acquisition.
\textbf{Rigoberto Hernandez:}
    Conceptualization,
    Writing -- Review \& Editing,
    Supervision,
    Project administration,
    Funding acquisition.


\section*{Acknowledgments}

Useful discussions with Robin Bardakcioglu are gratefully acknowledged.
The German portion of this collaborative work was partially supported
by the Deutsche Forschungsgemeinschaft (DFG) through Grant No.~MA1639/14-1.
The US portion was partially supported
by the National Science Foundation (NSF) through Grant No.~CHE 2102455.
This collaboration has also benefited from support
by the European Union's Horizon 2020 Research and Innovation Program
under the Marie Skłodowska-Curie Grant Agreement No.~734557.


\bibliographystyle{elsarticle-num-names}
\bibliography{paper-q34}

\begin{thebibliography}{68}
\expandafter\ifx\csname natexlab\endcsname\relax\def\natexlab#1{#1}\fi
\providecommand{\url}[1]{\texttt{#1}}
\providecommand{\href}[2]{#2}
\providecommand{\path}[1]{#1}
\providecommand{\DOIprefix}{doi:}
\providecommand{\ArXivprefix}{arXiv:}
\providecommand{\URLprefix}{URL: }
\providecommand{\Pubmedprefix}{pmid:}
\providecommand{\doi}[1]{\href{http://dx.doi.org/#1}{\path{#1}}}
\providecommand{\Pubmed}[1]{\href{pmid:#1}{\path{#1}}}
\providecommand{\bibinfo}[2]{#2}
\ifx\xfnm\relax \def\xfnm[#1]{\unskip,\space#1}\fi
\bibitem[{Augustine et~al.(2012)Augustine, Mojumder, Fong, Choday, Park, and
  Roy}]{Augustine2012a}
\bibinfo{author}{C.~Augustine}, \bibinfo{author}{N.~N. Mojumder},
  \bibinfo{author}{X.~Fong}, \bibinfo{author}{S.~H. Choday},
  \bibinfo{author}{S.~P. Park}, \bibinfo{author}{K.~Roy},
\newblock \bibinfo{title}{Spin-transfer torque {MRAMs} for low power memories:
  Perspective and prospective},
\newblock \bibinfo{journal}{IEEE Sens. J.} \bibinfo{volume}{12}
  (\bibinfo{year}{2012}) \bibinfo{pages}{756--766}.
  \DOIprefix\doi{10.1109/JSEN.2011.2124453}.
\bibitem[{Stoner and Wohlfarth(1948)}]{Stoner1948a}
\bibinfo{author}{E.~C. Stoner}, \bibinfo{author}{E.~P. Wohlfarth},
\newblock \bibinfo{title}{A mechanism of magnetic hysteresis in heterogeneous
  alloys},
\newblock \bibinfo{journal}{Philos. Trans. R. Soc., A} \bibinfo{volume}{240}
  (\bibinfo{year}{1948}) \bibinfo{pages}{599--642}.
  \DOIprefix\doi{10.1098/rsta.1948.0007}.
\bibitem[{Tannous and Gieraltowski(2008{\natexlab{a}})}]{Tannous2008a}
\bibinfo{author}{C.~Tannous}, \bibinfo{author}{J.~Gieraltowski},
\newblock \bibinfo{title}{The {Stoner}--{Wohlfarth} model of ferromagnetism},
\newblock \bibinfo{journal}{Eur. J. Phys.} \bibinfo{volume}{29}
  (\bibinfo{year}{2008}{\natexlab{a}}) \bibinfo{pages}{475--487}.
  \DOIprefix\doi{10.1088/0143-0807/29/3/008}.
\bibitem[{Tannous and Gieraltowski(2008{\natexlab{b}})}]{Tannous2008b}
\bibinfo{author}{C.~Tannous}, \bibinfo{author}{J.~Gieraltowski},
\newblock \bibinfo{title}{A {Stoner}--{Wohlfarth} model {Redux}: Static
  properties},
\newblock \bibinfo{journal}{Physica B} \bibinfo{volume}{403}
  (\bibinfo{year}{2008}{\natexlab{b}}) \bibinfo{pages}{3563--3570}.
  \DOIprefix\doi{10.1016/j.physb.2008.05.031}.
\bibitem[{Thirion et~al.(2003)Thirion, Wernsdorfer, and Mailly}]{Thirion2003a}
\bibinfo{author}{C.~Thirion}, \bibinfo{author}{W.~Wernsdorfer},
  \bibinfo{author}{D.~Mailly},
\newblock \bibinfo{title}{Switching of magnetization by nonlinear resonance
  studied in single nanoparticles},
\newblock \bibinfo{journal}{Nat. Mater.} \bibinfo{volume}{2}
  (\bibinfo{year}{2003}) \bibinfo{pages}{524--527}.
  \DOIprefix\doi{10.1038/nmat946}.
\bibitem[{Zhu et~al.(2008)Zhu, Zhu, and Tang}]{Zhu2008a}
\bibinfo{author}{J.-G. Zhu}, \bibinfo{author}{X.~Zhu},
  \bibinfo{author}{Y.~Tang},
\newblock \bibinfo{title}{Microwave assisted magnetic recording},
\newblock \bibinfo{journal}{IEEE Trans. Magn.} \bibinfo{volume}{44}
  (\bibinfo{year}{2008}) \bibinfo{pages}{125--131}.
  \DOIprefix\doi{10.1109/tmag.2007.911031}.
\bibitem[{Okamoto et~al.(2012)Okamoto, Kikuchi, Furuta, Kitakami, and
  Shimatsu}]{Okamoto2012a}
\bibinfo{author}{S.~Okamoto}, \bibinfo{author}{N.~Kikuchi},
  \bibinfo{author}{M.~Furuta}, \bibinfo{author}{O.~Kitakami},
  \bibinfo{author}{T.~Shimatsu},
\newblock \bibinfo{title}{Switching behaviors and its dynamics of a {Co}/{Pt}
  nanodot under the assistance of rf fields},
\newblock \bibinfo{journal}{Phys. Rev. Lett.} \bibinfo{volume}{109}
  (\bibinfo{year}{2012}) \bibinfo{pages}{237209}.
  \DOIprefix\doi{10.1103/physrevlett.109.237209}.
\bibitem[{Taniguchi(2014)}]{Taniguchi2014b}
\bibinfo{author}{T.~Taniguchi},
\newblock \bibinfo{title}{Magnetization reversal condition for a nanomagnet
  within a rotating magnetic field},
\newblock \bibinfo{journal}{Phys. Rev. B} \bibinfo{volume}{90}
  (\bibinfo{year}{2014}) \bibinfo{pages}{024424}.
  \DOIprefix\doi{10.1103/physrevb.90.024424}.
\bibitem[{Suto et~al.(2015)Suto, Nagasawa, Kudo, Mizushima, and
  Sato}]{Suto2015a}
\bibinfo{author}{H.~Suto}, \bibinfo{author}{T.~Nagasawa},
  \bibinfo{author}{K.~Kudo}, \bibinfo{author}{K.~Mizushima},
  \bibinfo{author}{R.~Sato},
\newblock \bibinfo{title}{Microwave-assisted switching of a single
  perpendicular magnetic tunnel junction nanodot},
\newblock \bibinfo{journal}{Appl. Phys. Express} \bibinfo{volume}{8}
  (\bibinfo{year}{2015}) \bibinfo{pages}{023001}.
  \DOIprefix\doi{10.7567/apex.8.023001}.
\bibitem[{Barros et~al.(2011)Barros, Rassam, Jirari, and
  Kachkachi}]{Barros2011a}
\bibinfo{author}{N.~Barros}, \bibinfo{author}{M.~Rassam},
  \bibinfo{author}{H.~Jirari}, \bibinfo{author}{H.~Kachkachi},
\newblock \bibinfo{title}{Optimal switching of a nanomagnet assisted by
  microwaves},
\newblock \bibinfo{journal}{Phys. Rev. B} \bibinfo{volume}{83}
  (\bibinfo{year}{2011}) \bibinfo{pages}{144418}.
  \DOIprefix\doi{10.1103/physrevb.83.144418}.
\bibitem[{Barros et~al.(2013)Barros, Rassam, and Kachkachi}]{Barros2013a}
\bibinfo{author}{N.~Barros}, \bibinfo{author}{H.~Rassam},
  \bibinfo{author}{H.~Kachkachi},
\newblock \bibinfo{title}{Microwave-assisted switching of a nanomagnet:
  Analytical determination of the optimal microwave field},
\newblock \bibinfo{journal}{Phys. Rev. B} \bibinfo{volume}{88}
  (\bibinfo{year}{2013}) \bibinfo{pages}{014421}.
  \DOIprefix\doi{10.1103/physrevb.88.014421}.
\bibitem[{Klughertz et~al.(2015)Klughertz, Friedland, Hervieux, and
  Manfredi}]{Klughertz2015a}
\bibinfo{author}{G.~Klughertz}, \bibinfo{author}{L.~Friedland},
  \bibinfo{author}{P.-A. Hervieux}, \bibinfo{author}{G.~Manfredi},
\newblock \bibinfo{title}{Autoresonant switching of the magnetization in
  single-domain nanoparticles: Two-level theory},
\newblock \bibinfo{journal}{Phys. Rev. B} \bibinfo{volume}{91}
  (\bibinfo{year}{2015}) \bibinfo{pages}{104433}.
  \DOIprefix\doi{10.1103/physrevb.91.104433}.
\bibitem[{Taniguchi et~al.(2016)Taniguchi, Saida, Nakatani, and
  Kubota}]{Taniguchi2016a}
\bibinfo{author}{T.~Taniguchi}, \bibinfo{author}{D.~Saida},
  \bibinfo{author}{Y.~Nakatani}, \bibinfo{author}{H.~Kubota},
\newblock \bibinfo{title}{Magnetization switching by current and microwaves},
\newblock \bibinfo{journal}{Phys. Rev. B} \bibinfo{volume}{93}
  (\bibinfo{year}{2016}) \bibinfo{pages}{014430}.
  \DOIprefix\doi{10.1103/physrevb.93.014430}.
\bibitem[{Rivkin and Ketterson(2006)}]{Rivkin2006a}
\bibinfo{author}{K.~Rivkin}, \bibinfo{author}{J.~B. Ketterson},
\newblock \bibinfo{title}{Magnetization reversal in the anisotropy-dominated
  regime using time-dependent magnetic fields},
\newblock \bibinfo{journal}{Appl. Phys. Lett.} \bibinfo{volume}{89}
  (\bibinfo{year}{2006}) \bibinfo{pages}{252507}.
  \DOIprefix\doi{10.1063/1.2405855}.
\bibitem[{Taniguchi(2015)}]{Taniguchi2015a}
\bibinfo{author}{T.~Taniguchi},
\newblock \bibinfo{title}{Magnetization switching by microwaves synchronized in
  the vicinity of precession frequency},
\newblock \bibinfo{journal}{Appl. Phys. Express} \bibinfo{volume}{8}
  (\bibinfo{year}{2015}) \bibinfo{pages}{083004}.
  \DOIprefix\doi{10.7567/apex.8.083004}.
\bibitem[{M{\"o}gerle et~al.(2022)M{\"o}gerle, Schuldt, Reiff, Main, and
  Hernandez}]{hern22b}
\bibinfo{author}{J.~M{\"o}gerle}, \bibinfo{author}{R.~Schuldt},
  \bibinfo{author}{J.~Reiff}, \bibinfo{author}{J.~Main},
  \bibinfo{author}{R.~Hernandez},
\newblock \bibinfo{title}{Transition state dynamics of a driven magnetic free
  layer},
\newblock \bibinfo{journal}{Commun. Nonlinear Sci. Numer. Simulat.}
  \bibinfo{volume}{105} (\bibinfo{year}{2022}) \bibinfo{pages}{106054}.
  \DOIprefix\doi{10.1016/j.cnsns.2021.106054}.
\bibitem[{Feldmaier et~al.(2019)Feldmaier, Schraft, Bardakcioglu, Reiff, Lober,
  Tsch{\"o}pe, Junginger, Main, Bartsch, and Hernandez}]{hern19a}
\bibinfo{author}{M.~Feldmaier}, \bibinfo{author}{P.~Schraft},
  \bibinfo{author}{R.~Bardakcioglu}, \bibinfo{author}{J.~Reiff},
  \bibinfo{author}{M.~Lober}, \bibinfo{author}{M.~Tsch{\"o}pe},
  \bibinfo{author}{A.~Junginger}, \bibinfo{author}{J.~Main},
  \bibinfo{author}{T.~Bartsch}, \bibinfo{author}{R.~Hernandez},
\newblock \bibinfo{title}{Invariant manifolds and rate constants in driven
  chemical reactions},
\newblock \bibinfo{journal}{J. Phys. Chem. B} \bibinfo{volume}{123}
  (\bibinfo{year}{2019}) \bibinfo{pages}{2070--2086}.
  \DOIprefix\doi{10.1021/acs.jpcb.8b10541}.
\bibitem[{Nagahata et~al.(2021)Nagahata, Hernandez, and Komatsuzaki}]{hern21j}
\bibinfo{author}{Y.~Nagahata}, \bibinfo{author}{R.~Hernandez},
  \bibinfo{author}{T.~Komatsuzaki},
\newblock \bibinfo{title}{Phase space geometry of isolated to condensed
  chemical reactions},
\newblock \bibinfo{journal}{J. Chem. Phys.} \bibinfo{volume}{155}
  (\bibinfo{year}{2021}) \bibinfo{pages}{210901}.
  \DOIprefix\doi{10.1063/5.0059618}.
\bibitem[{H{\"a}nggi et~al.(1990)H{\"a}nggi, Talkner, and Borkovec}]{rmp90}
\bibinfo{author}{P.~H{\"a}nggi}, \bibinfo{author}{P.~Talkner},
  \bibinfo{author}{M.~Borkovec},
\newblock \bibinfo{title}{Reaction-rate theory: Fifty years after {Kramers}},
\newblock \bibinfo{journal}{Rev. Mod. Phys.} \bibinfo{volume}{62}
  (\bibinfo{year}{1990}) \bibinfo{pages}{251--341}.
  \DOIprefix\doi{10.1103/RevModPhys.62.251}, \bibinfo{note}{and references
  therein}.
\bibitem[{Jaff{\'e} et~al.(2000)Jaff{\'e}, Farrelly, and Uzer}]{Jaffe00}
\bibinfo{author}{C.~Jaff{\'e}}, \bibinfo{author}{D.~Farrelly},
  \bibinfo{author}{T.~Uzer},
\newblock \bibinfo{title}{Transition state theory without time-reversal
  symmetry: Chaotic ionization of the hydrogen atom},
\newblock \bibinfo{journal}{Phys. Rev. Lett.} \bibinfo{volume}{84}
  (\bibinfo{year}{2000}) \bibinfo{pages}{610--613}.
  \DOIprefix\doi{10.1103/PhysRevLett.84.610}.
\bibitem[{Jacucci et~al.(1984)Jacucci, Toller, DeLorenzi, and
  Flynn}]{Jacucci1984}
\bibinfo{author}{G.~Jacucci}, \bibinfo{author}{M.~Toller},
  \bibinfo{author}{G.~DeLorenzi}, \bibinfo{author}{C.~P. Flynn},
\newblock \bibinfo{title}{Rate theory, return jump catastrophes, and the center
  manifold},
\newblock \bibinfo{journal}{Phys. Rev. Lett.} \bibinfo{volume}{52}
  (\bibinfo{year}{1984}) \bibinfo{pages}{295}.
  \DOIprefix\doi{10.1103/PhysRevLett.52.295}.
\bibitem[{Komatsuzaki and Berry(1999)}]{KomatsuzakiBerry99a}
\bibinfo{author}{T.~Komatsuzaki}, \bibinfo{author}{R.~S. Berry},
\newblock \bibinfo{title}{Regularity in chaotic reaction paths. {I}. ${\rm
  ar}_6$},
\newblock \bibinfo{journal}{J. Chem. Phys.} \bibinfo{volume}{110}
  (\bibinfo{year}{1999}) \bibinfo{pages}{9160--9173}.
  \DOIprefix\doi{10.1063/1.478838}.
\bibitem[{Komatsuzaki and Berry(2002)}]{KomatsuzakiBerry02}
\bibinfo{author}{T.~Komatsuzaki}, \bibinfo{author}{R.~S. Berry},
\newblock \bibinfo{title}{Chemical reaction dynamics: Many-body chaos and
  regularity},
\newblock \bibinfo{journal}{Adv. Chem. Phys.} \bibinfo{volume}{123}
  (\bibinfo{year}{2002}) \bibinfo{pages}{79--152}.
  \DOIprefix\doi{10.1002/0471231509.ch2}.
\bibitem[{Toller et~al.(1985)Toller, Jacucci, DeLorenzi, and Flynn}]{toller}
\bibinfo{author}{M.~Toller}, \bibinfo{author}{G.~Jacucci},
  \bibinfo{author}{G.~DeLorenzi}, \bibinfo{author}{C.~P. Flynn},
\newblock \bibinfo{title}{Theory of classical diffusion jumps in solids},
\newblock \bibinfo{journal}{Phys. Rev. B} \bibinfo{volume}{32}
  (\bibinfo{year}{1985}) \bibinfo{pages}{2082}.
  \DOIprefix\doi{10.1103/PhysRevB.32.2082}.
\bibitem[{Voter et~al.(2002)Voter, Montalenti, and Germann}]{voter02b}
\bibinfo{author}{A.~F. Voter}, \bibinfo{author}{F.~Montalenti},
  \bibinfo{author}{T.~C. Germann},
\newblock \bibinfo{title}{Extending the time scale in atomistic simulations of
  materials},
\newblock \bibinfo{journal}{Annu.~Rev.~Mater.~Res.} \bibinfo{volume}{32}
  (\bibinfo{year}{2002}) \bibinfo{pages}{321--346}.
  \DOIprefix\doi{10.1146/annurev.matsci.32.112601.141541}.
\bibitem[{de~Oliveira et~al.(2002)de~Oliveira, {Ozorio de Almeida}, {Dami{\~a}o
  Soares}, and Tonini}]{Oliveira02}
\bibinfo{author}{H.~P. de~Oliveira}, \bibinfo{author}{A.~M. {Ozorio de
  Almeida}}, \bibinfo{author}{I.~{Dami{\~a}o Soares}}, \bibinfo{author}{E.~V.
  Tonini},
\newblock \bibinfo{title}{Homoclinic chaos in the dynamics of a general
  {Bianchi} type-{IX} model},
\newblock \bibinfo{journal}{Phys. Rev. D} \bibinfo{volume}{65}
  (\bibinfo{year}{2002}) \bibinfo{pages}{083511/1--9}.
  \DOIprefix\doi{10.1103/PhysRevD.65.083511}.
\bibitem[{Jaff{\'e} et~al.(2002)Jaff{\'e}, Ross, Lo, Marsden, Farrelly, and
  Uzer}]{Jaffe02}
\bibinfo{author}{C.~Jaff{\'e}}, \bibinfo{author}{S.~D. Ross},
  \bibinfo{author}{M.~W. Lo}, \bibinfo{author}{J.~Marsden},
  \bibinfo{author}{D.~Farrelly}, \bibinfo{author}{T.~Uzer},
\newblock \bibinfo{title}{Statistical theory of asteroid escape rates},
\newblock \bibinfo{journal}{Phys. Rev. Lett.} \bibinfo{volume}{89}
  (\bibinfo{year}{2002}) \bibinfo{pages}{011101}.
  \DOIprefix\doi{10.1103/PhysRevLett.89.011101}.
\bibitem[{Reiff et~al.(2022)Reiff, Zatsch, Main, and Hernandez}]{hern22a}
\bibinfo{author}{J.~Reiff}, \bibinfo{author}{J.~Zatsch},
  \bibinfo{author}{J.~Main}, \bibinfo{author}{R.~Hernandez},
\newblock \bibinfo{title}{On the stability of satellites at unstable libration
  points of sun--planet--moon systems},
\newblock \bibinfo{journal}{Commun. Nonlinear Sci. Numer. Simulat.}
  \bibinfo{volume}{104} (\bibinfo{year}{2022}) \bibinfo{pages}{106053}.
  \DOIprefix\doi{10.1016/j.cnsns.2021.106053}.
\bibitem[{Bessarab et~al.(2012)Bessarab, Uzdin, and
  J{\'o}nsson}]{Bessarab2012a}
\bibinfo{author}{P.~F. Bessarab}, \bibinfo{author}{V.~M. Uzdin},
  \bibinfo{author}{H.~J{\'o}nsson},
\newblock \bibinfo{title}{Harmonic transition-state theory of thermal spin
  transitions},
\newblock \bibinfo{journal}{Phys. Rev. B} \bibinfo{volume}{85}
  (\bibinfo{year}{2012}) \bibinfo{pages}{184409}.
  \DOIprefix\doi{10.1103/PhysRevB.85.184409}.
\bibitem[{Wang and Visscher(2007)}]{Wang2007a}
\bibinfo{author}{S.~Wang}, \bibinfo{author}{P.~B. Visscher},
\newblock \bibinfo{title}{Accelerated {LLG} simulation of magnetic stability:
  ``bounce'' algorithm},
\newblock \bibinfo{journal}{IEEE Trans. Magn.} \bibinfo{volume}{43}
  (\bibinfo{year}{2007}) \bibinfo{pages}{2893--2895}.
  \DOIprefix\doi{10.1109/tmag.2007.892595}.
\bibitem[{Ciornei et~al.(2011)Ciornei, Rubi, and Wegrowe}]{Ciornei2011}
\bibinfo{author}{M.-C. Ciornei}, \bibinfo{author}{J.~M. Rubi},
  \bibinfo{author}{J.-E. Wegrowe},
\newblock \bibinfo{title}{Magnetization dynamics in the inertial regime:
  Nutation predicted at short time scales},
\newblock \bibinfo{journal}{Phys. Rev. B} \bibinfo{volume}{83}
  (\bibinfo{year}{2011}) \bibinfo{pages}{020410(R)}.
  \DOIprefix\doi{10.1103/PhysRevB.83.020410}.
\bibitem[{F{\"a}hnle et~al.(2011)F{\"a}hnle, Steiauf, and Illg}]{Fahnle2011}
\bibinfo{author}{M.~F{\"a}hnle}, \bibinfo{author}{D.~Steiauf},
  \bibinfo{author}{C.~Illg},
\newblock \bibinfo{title}{Generalized {Gilbert} equation including inertial
  damping: Derivation from an extended breathing {Fermi} surface model},
\newblock \bibinfo{journal}{Phys. Rev. B} \bibinfo{volume}{84}
  (\bibinfo{year}{2011}) \bibinfo{pages}{172403}.
  \DOIprefix\doi{10.1103/PhysRevB.84.172403}.
\bibitem[{Li et~al.(2015)Li, Barra, Auffret, Ebels, and Bailey}]{LiBarra2015}
\bibinfo{author}{Y.~Li}, \bibinfo{author}{A.-L. Barra},
  \bibinfo{author}{S.~Auffret}, \bibinfo{author}{U.~Ebels},
  \bibinfo{author}{W.~E. Bailey},
\newblock \bibinfo{title}{Inertial terms to magnetization dynamics in
  ferromagnetic thin films},
\newblock \bibinfo{journal}{Phys. Rev. B} \bibinfo{volume}{92}
  (\bibinfo{year}{2015}) \bibinfo{pages}{140413(R)}.
  \DOIprefix\doi{10.1103/PhysRevB.92.140413}.
\bibitem[{Neeraj et~al.(2021)Neeraj, Awari, Kovalev, Polley, Hagström,
  Arekapudi, Semisalova, Lenz, Green, Deinert, Ilyakov, Chen, Bawatna, Scalera,
  d'Aquino, Serpico, Hellwig, Wegrowe, Gensch, and Bonetti}]{Neeraj2021a}
\bibinfo{author}{K.~Neeraj}, \bibinfo{author}{N.~Awari},
  \bibinfo{author}{S.~Kovalev}, \bibinfo{author}{D.~Polley},
  \bibinfo{author}{N.~Z. Hagström}, \bibinfo{author}{S.~S. P.~K. Arekapudi},
  \bibinfo{author}{A.~Semisalova}, \bibinfo{author}{K.~Lenz},
  \bibinfo{author}{B.~Green}, \bibinfo{author}{J.-C. Deinert},
  \bibinfo{author}{I.~Ilyakov}, \bibinfo{author}{M.~Chen},
  \bibinfo{author}{M.~Bawatna}, \bibinfo{author}{V.~Scalera},
  \bibinfo{author}{M.~d'Aquino}, \bibinfo{author}{C.~Serpico},
  \bibinfo{author}{O.~Hellwig}, \bibinfo{author}{J.-E. Wegrowe},
  \bibinfo{author}{M.~Gensch}, \bibinfo{author}{S.~Bonetti},
\newblock \bibinfo{title}{Inertial spin dynamics in ferromagnets},
\newblock \bibinfo{journal}{Nat. Phys.} \bibinfo{volume}{17}
  (\bibinfo{year}{2021}) \bibinfo{pages}{245--250}.
  \DOIprefix\doi{10.1038/s41567-020-01040-y}.
\bibitem[{Gilbert(2004)}]{Gilbert2004}
\bibinfo{author}{T.~L. Gilbert},
\newblock \bibinfo{title}{A phenomenological theory of damping in ferromagnetic
  materials},
\newblock \bibinfo{journal}{IEEE Trans. Magn.} \bibinfo{volume}{40}
  (\bibinfo{year}{2004}) \bibinfo{pages}{3443--3449}.
  \DOIprefix\doi{10.1109/TMAG.2004.836740}.
\bibitem[{Apalkov and Visscher(2005)}]{Apalkov2005}
\bibinfo{author}{D.~M. Apalkov}, \bibinfo{author}{P.~B. Visscher},
\newblock \bibinfo{title}{Spin-torque switching: {Fokker-Planck} rate
  calculation},
\newblock \bibinfo{journal}{Phys. Rev. B} \bibinfo{volume}{72}
  (\bibinfo{year}{2005}) \bibinfo{pages}{180405(R)}.
  \DOIprefix\doi{10.1103/PhysRevB.72.180405}.
\bibitem[{Abert(2019)}]{Abert2019a}
\bibinfo{author}{C.~Abert},
\newblock \bibinfo{title}{Micromagnetics and spintronics: Models and numerical
  methods},
\newblock \bibinfo{journal}{Eur. Phys. J. B} \bibinfo{volume}{92}
  (\bibinfo{year}{2019}) \bibinfo{pages}{120}.
  \DOIprefix\doi{10.1140/epjb/e2019-90599-6}.
\bibitem[{Steiauf and F{\"a}hnle(2005)}]{Steiauf2005a}
\bibinfo{author}{D.~Steiauf}, \bibinfo{author}{M.~F{\"a}hnle},
\newblock \bibinfo{title}{Damping of spin dynamics in nanostructures: An ab
  initio study},
\newblock \bibinfo{journal}{Phys. Rev. B} \bibinfo{volume}{72}
  (\bibinfo{year}{2005}) \bibinfo{pages}{064450}.
  \DOIprefix\doi{10.1103/PhysRevB.72.064450}.
\bibitem[{F{\"a}hnle and Steiauf(2006)}]{Faehnle2006a}
\bibinfo{author}{M.~F{\"a}hnle}, \bibinfo{author}{D.~Steiauf},
\newblock \bibinfo{title}{Breathing {Fermi} surface model for noncollinear
  magnetization: A generalization of the {Gilbert} equation},
\newblock \bibinfo{journal}{Phys. Rev. B} \bibinfo{volume}{73}
  (\bibinfo{year}{2006}) \bibinfo{pages}{184427}.
  \DOIprefix\doi{10.1103/PhysRevB.73.184427}.
\bibitem[{Bauer et~al.(2000)Bauer, Fassbender, Hillebrands, and
  Stamps}]{Bauer2000a}
\bibinfo{author}{M.~Bauer}, \bibinfo{author}{J.~Fassbender},
  \bibinfo{author}{B.~Hillebrands}, \bibinfo{author}{R.~L. Stamps},
\newblock \bibinfo{title}{Switching behavior of a {Stoner} particle beyond the
  relaxation time limit},
\newblock \bibinfo{journal}{Phys. Rev. B} \bibinfo{volume}{61}
  (\bibinfo{year}{2000}) \bibinfo{pages}{3410--3416}.
  \DOIprefix\doi{10.1103/PhysRevB.61.3410}.
\bibitem[{Sun(2000)}]{Sun2000a}
\bibinfo{author}{J.~Z. Sun},
\newblock \bibinfo{title}{Spin-current interaction with a monodomain magnetic
  body: A model study},
\newblock \bibinfo{journal}{Phys. Rev. B} \bibinfo{volume}{62}
  (\bibinfo{year}{2000}) \bibinfo{pages}{570--578}.
  \DOIprefix\doi{10.1103/PhysRevB.62.570}.
\bibitem[{Wegrowe and Ciornei(2012)}]{Wegrowe2012a}
\bibinfo{author}{J.-E. Wegrowe}, \bibinfo{author}{M.-C. Ciornei},
\newblock \bibinfo{title}{Magnetization dynamics, gyromagnetic relation, and
  inertial effects},
\newblock \bibinfo{journal}{Am. J. Phys.} \bibinfo{volume}{80}
  (\bibinfo{year}{2012}) \bibinfo{pages}{607--611}.
  \DOIprefix\doi{10.1119/1.4709188}.
\bibitem[{Kikuchi and Tatara(2015)}]{Kikuchi2015a}
\bibinfo{author}{T.~Kikuchi}, \bibinfo{author}{G.~Tatara},
\newblock \bibinfo{title}{Spin dynamics with inertia in metallic ferromagnets},
\newblock \bibinfo{journal}{Phys. Rev. B} \bibinfo{volume}{92}
  (\bibinfo{year}{2015}) \bibinfo{pages}{184410}.
  \DOIprefix\doi{10.1103/PhysRevB.92.184410}.
\bibitem[{Garrett and Truhlar(1979)}]{truh79}
\bibinfo{author}{B.~C. Garrett}, \bibinfo{author}{D.~G. Truhlar},
\newblock \bibinfo{title}{Generalized transition state theory},
\newblock \bibinfo{journal}{J. Phys. Chem.} \bibinfo{volume}{83}
  (\bibinfo{year}{1979}) \bibinfo{pages}{1052--1079}.
  \DOIprefix\doi{10.1021/j100471a031}.
\bibitem[{Pechukas(1981)}]{pech81}
\bibinfo{author}{P.~Pechukas},
\newblock \bibinfo{title}{Transition state theory},
\newblock \bibinfo{journal}{Annu. Rev. Phys. Chem.} \bibinfo{volume}{32}
  (\bibinfo{year}{1981}) \bibinfo{pages}{159--177}.
  \DOIprefix\doi{10.1146/annurev.pc.32.100181.001111}.
\bibitem[{Pollak(1985)}]{pollak85}
\bibinfo{author}{E.~Pollak},
\newblock \bibinfo{title}{Periodic orbits and the theory of reactive
  scattering},
\newblock in: \bibinfo{editor}{M.~Baer} (Ed.), \bibinfo{booktitle}{Theory of
  Chemical Reaction Dynamics}, volume~\bibinfo{volume}{3},
  \bibinfo{publisher}{CRC Press}, \bibinfo{address}{Boca Raton, FL},
  \bibinfo{year}{1985}, p. \bibinfo{pages}{123}.
\bibitem[{Truhlar et~al.(1996)Truhlar, Garrett, and Klippenstein}]{truh96}
\bibinfo{author}{D.~G. Truhlar}, \bibinfo{author}{B.~C. Garrett},
  \bibinfo{author}{S.~J. Klippenstein},
\newblock \bibinfo{title}{Current status of transition-state theory},
\newblock \bibinfo{journal}{J. Phys. Chem.} \bibinfo{volume}{100}
  (\bibinfo{year}{1996}) \bibinfo{pages}{12771--12800}.
  \DOIprefix\doi{10.1021/jp953748q}.
\bibitem[{Miller(1998)}]{mill98}
\bibinfo{author}{W.~H. Miller},
\newblock \bibinfo{title}{Direct and correct calculation of canonical and
  microcanonical rate constants for chemical reactions},
\newblock \bibinfo{journal}{J. Phys. Chem. A} \bibinfo{volume}{102}
  (\bibinfo{year}{1998}) \bibinfo{pages}{793--806}.
  \DOIprefix\doi{10.1021/jp973208o}.
\bibitem[{Hernandez et~al.(2010)Hernandez, Uzer, and Bartsch}]{hern10a}
\bibinfo{author}{R.~Hernandez}, \bibinfo{author}{T.~Uzer},
  \bibinfo{author}{T.~Bartsch},
\newblock \bibinfo{title}{Transition state theory in liquids beyond planar
  dividing surfaces},
\newblock \bibinfo{journal}{Chem. Phys.} \bibinfo{volume}{370}
  (\bibinfo{year}{2010}) \bibinfo{pages}{270--276}.
  \DOIprefix\doi{10.1016/j.chemphys.2010.01.016}.
\bibitem[{Hernandez et~al.(2017)Hernandez, Stallings, and Iyer}]{hern17a}
\bibinfo{author}{R.~Hernandez}, \bibinfo{author}{D.~Stallings},
  \bibinfo{author}{S.~Iyer},
\newblock \bibinfo{title}{The gender and urm faculty demographics data
  collected by oxide},
\newblock in: \bibinfo{editor}{H.~N. Cheng}, \bibinfo{editor}{D.~Nelson}
  (Eds.), \bibinfo{booktitle}{Diversity in the Scientific Community Volume 1:
  Quantifying Diversity and Formulating Success}, volume \bibinfo{volume}{1255}
  of \textit{\bibinfo{series}{ACS Symposium Series}},
  \bibinfo{publisher}{American Chemical Society; Oxford University Press},
  \bibinfo{address}{Washington DC}, \bibinfo{year}{2017}, pp.
  \bibinfo{pages}{101--112}. \DOIprefix\doi{10.1021/bk-2017-1256.ch006}.
\bibitem[{Feldmaier et~al.(2019)Feldmaier, Bardakcioglu, Reiff, Main, and
  Hernandez}]{hern19e}
\bibinfo{author}{M.~Feldmaier}, \bibinfo{author}{R.~Bardakcioglu},
  \bibinfo{author}{J.~Reiff}, \bibinfo{author}{J.~Main},
  \bibinfo{author}{R.~Hernandez},
\newblock \bibinfo{title}{Phase-space resolved rates in driven multidimensional
  chemical reactions},
\newblock \bibinfo{journal}{J. Chem. Phys.} \bibinfo{volume}{151}
  (\bibinfo{year}{2019}) \bibinfo{pages}{244108}.
  \DOIprefix\doi{10.1063/1.5127539}.
\bibitem[{Feldmaier et~al.(2020)Feldmaier, Reiff, Benito, Borondo, Main, and
  Hernandez}]{hern20m}
\bibinfo{author}{M.~Feldmaier}, \bibinfo{author}{J.~Reiff},
  \bibinfo{author}{R.~M. Benito}, \bibinfo{author}{F.~Borondo},
  \bibinfo{author}{J.~Main}, \bibinfo{author}{R.~Hernandez},
\newblock \bibinfo{title}{Influence of external driving on decays in the
  geometry of the {LiCN} isomerization},
\newblock \bibinfo{journal}{J. Chem. Phys.} \bibinfo{volume}{153}
  (\bibinfo{year}{2020}) \bibinfo{pages}{084115}.
  \DOIprefix\doi{10.1063/5.0015509}.
\bibitem[{Bardakcioglu et~al.(2020)Bardakcioglu, Reiff, Feldmaier, Main, and
  Hernandez}]{hern20o}
\bibinfo{author}{R.~Bardakcioglu}, \bibinfo{author}{J.~Reiff},
  \bibinfo{author}{M.~Feldmaier}, \bibinfo{author}{J.~Main},
  \bibinfo{author}{R.~Hernandez},
\newblock \bibinfo{title}{Thermal decay rates of an activated complex in a
  driven model chemical reaction},
\newblock \bibinfo{journal}{Phys. Rev. E} \bibinfo{volume}{102}
  (\bibinfo{year}{2020}) \bibinfo{pages}{062204}.
  \DOIprefix\doi{10.1103/PhysRevE.102.062204}.
\bibitem[{Lichtenberg and Liebermann(1982)}]{Lichtenberg82}
\bibinfo{author}{A.~J. Lichtenberg}, \bibinfo{author}{M.~A. Liebermann},
  \bibinfo{title}{Regular and Stochastic Motion},
  \bibinfo{publisher}{Springer}, \bibinfo{address}{New York},
  \bibinfo{year}{1982}.
\bibitem[{Hernandez and Miller(1993)}]{hern93b}
\bibinfo{author}{R.~Hernandez}, \bibinfo{author}{W.~H. Miller},
\newblock \bibinfo{title}{Semiclassical transition state theory. {A} new
  perspective},
\newblock \bibinfo{journal}{Chem. Phys. Lett.} \bibinfo{volume}{214}
  (\bibinfo{year}{1993}) \bibinfo{pages}{129--136}.
  \DOIprefix\doi{10.1016/0009-2614(93)90071-8}.
\bibitem[{Hernandez(1993)}]{hern93c}
\bibinfo{author}{R.~Hernandez}, \bibinfo{title}{Application of Semiclassical
  Methods to Reaction Rate Theory}, Ph.D. thesis, University of California,
  \bibinfo{address}{Berkeley, CA}, \bibinfo{year}{1993}.
\bibitem[{Wiggins et~al.(2001)Wiggins, Wiesenfeld, Jaffe, and Uzer}]{wiggins01}
\bibinfo{author}{S.~Wiggins}, \bibinfo{author}{L.~Wiesenfeld},
  \bibinfo{author}{C.~Jaffe}, \bibinfo{author}{T.~Uzer},
\newblock \bibinfo{title}{Impenetrable barriers in phase-space},
\newblock \bibinfo{journal}{Phys. Rev. Lett.} \bibinfo{volume}{86}
  (\bibinfo{year}{2001}) \bibinfo{pages}{5478}.
  \DOIprefix\doi{10.1103/PhysRevLett.86.5478}.
\bibitem[{Uzer et~al.(2002)Uzer, Jaff\'e, Palaci{\'a}n, Yanguas, and
  Wiggins}]{Uzer02}
\bibinfo{author}{T.~Uzer}, \bibinfo{author}{C.~Jaff\'e},
  \bibinfo{author}{J.~Palaci{\'a}n}, \bibinfo{author}{P.~Yanguas},
  \bibinfo{author}{S.~Wiggins},
\newblock \bibinfo{title}{The geometry of reaction dynamics},
\newblock \bibinfo{journal}{Nonlinearity} \bibinfo{volume}{15}
  (\bibinfo{year}{2002}) \bibinfo{pages}{957--992}.
  \DOIprefix\doi{10.1088/0951-7715/15/4/301}.
\bibitem[{Ott(2002)}]{Ott2002a}
\bibinfo{author}{E.~Ott}, \bibinfo{title}{Chaos in Dynamical Systems},
  \bibinfo{edition}{2nd} ed., \bibinfo{publisher}{Cambridge University Press},
  \bibinfo{address}{Cambridge, England}, \bibinfo{year}{2002}.
\bibitem[{Wiggins(2016)}]{wiggins16}
\bibinfo{author}{S.~Wiggins},
\newblock \bibinfo{title}{The role of normally hyperbolic invariant manifolds
  ({NHIMS}) in the context of the phase space setting for chemical reaction
  dynamics},
\newblock \bibinfo{journal}{Regul. Chaotic Dyn.} \bibinfo{volume}{21}
  (\bibinfo{year}{2016}) \bibinfo{pages}{621--638}.
  \DOIprefix\doi{10.1134/S1560354716060034}.
\bibitem[{Mendoza and Mancho(2010)}]{Mancho2010}
\bibinfo{author}{C.~Mendoza}, \bibinfo{author}{A.~M. Mancho},
\newblock \bibinfo{title}{Hidden geometry of ocean flows},
\newblock \bibinfo{journal}{Phys. Rev. Lett.} \bibinfo{volume}{105}
  (\bibinfo{year}{2010}) \bibinfo{pages}{038501}.
  \DOIprefix\doi{10.1103/PhysRevLett.105.038501}.
\bibitem[{Mancho et~al.(2013)Mancho, Wiggins, Curbelo, and
  Mendoza}]{Mancho2013}
\bibinfo{author}{A.~M. Mancho}, \bibinfo{author}{S.~Wiggins},
  \bibinfo{author}{J.~Curbelo}, \bibinfo{author}{C.~Mendoza},
\newblock \bibinfo{title}{{Lagrangian} descriptors: A method for revealing
  phase space structures of general time dependent dynamical systems},
\newblock \bibinfo{journal}{Commun. Nonlinear Sci. Numer. Simul.}
  \bibinfo{volume}{18} (\bibinfo{year}{2013}) \bibinfo{pages}{3530 -- 3557}.
  \DOIprefix\doi{10.1016/j.cnsns.2013.05.002}.
\bibitem[{Craven and Hernandez(2015)}]{hern15e}
\bibinfo{author}{G.~T. Craven}, \bibinfo{author}{R.~Hernandez},
\newblock \bibinfo{title}{{Lagrangian} descriptors of thermalized transition
  states on time-varying energy surfaces},
\newblock \bibinfo{journal}{Phys. Rev. Lett.} \bibinfo{volume}{115}
  (\bibinfo{year}{2015}) \bibinfo{pages}{148301}.
  \DOIprefix\doi{10.1103/PhysRevLett.115.148301}.
\bibitem[{Reiff et~al.(2021)Reiff, Feldmaier, Main, and Hernandez}]{hern21a}
\bibinfo{author}{J.~Reiff}, \bibinfo{author}{M.~Feldmaier},
  \bibinfo{author}{J.~Main}, \bibinfo{author}{R.~Hernandez},
\newblock \bibinfo{title}{Dynamics and decay rates of a time-dependent
  two-saddle system},
\newblock \bibinfo{journal}{Phys. Rev. E} \bibinfo{volume}{103}
  (\bibinfo{year}{2021}) \bibinfo{pages}{022121}.
  \DOIprefix\doi{10.1103/PhysRevE.103.022121}.
\bibitem[{Craven et~al.(2014)Craven, Bartsch, and Hernandez}]{hern14f}
\bibinfo{author}{G.~T. Craven}, \bibinfo{author}{T.~Bartsch},
  \bibinfo{author}{R.~Hernandez},
\newblock \bibinfo{title}{Communication: Transition state trajectory stability
  determines barrier crossing rates in chemical reactions induced by
  time-dependent oscillating fields},
\newblock \bibinfo{journal}{J. Chem. Phys.} \bibinfo{volume}{141}
  (\bibinfo{year}{2014}) \bibinfo{pages}{041106}.
  \DOIprefix\doi{10.1063/1.4891471}.
\bibitem[{Bardakcioglu et~al.(2018)Bardakcioglu, Junginger, Feldmaier, Main,
  and Hernandez}]{hern18g}
\bibinfo{author}{R.~Bardakcioglu}, \bibinfo{author}{A.~Junginger},
  \bibinfo{author}{M.~Feldmaier}, \bibinfo{author}{J.~Main},
  \bibinfo{author}{R.~Hernandez},
\newblock \bibinfo{title}{Binary contraction method for the construction of
  time-dependent dividing surfaces in driven chemical reactions},
\newblock \bibinfo{journal}{Phys. Rev. E} \bibinfo{volume}{98}
  (\bibinfo{year}{2018}) \bibinfo{pages}{032204}.
  \DOIprefix\doi{10.1103/PhysRevE.98.032204}.
\bibitem[{Olive et~al.(2015)Olive, Lansac, Meyer, Hayoun, and
  Wegrowe}]{Olive2015a}
\bibinfo{author}{E.~Olive}, \bibinfo{author}{Y.~Lansac},
  \bibinfo{author}{M.~Meyer}, \bibinfo{author}{M.~Hayoun},
  \bibinfo{author}{J.-E. Wegrowe},
\newblock \bibinfo{title}{Deviation from the {Landau-Lifshitz-Gilbert} equation
  in the inertial regime of the magnetization},
\newblock \bibinfo{journal}{J. Appl. Phys.} \bibinfo{volume}{117}
  (\bibinfo{year}{2015}) \bibinfo{pages}{213904}.
  \DOIprefix\doi{10.1063/1.4921908}.
\bibitem[{Reiff et~al.(2021)Reiff, Bardakcioglu, Feldmaier, Main, and
  Hernandez}]{hern21h}
\bibinfo{author}{J.~Reiff}, \bibinfo{author}{R.~Bardakcioglu},
  \bibinfo{author}{M.~Feldmaier}, \bibinfo{author}{J.~Main},
  \bibinfo{author}{R.~Hernandez},
\newblock \bibinfo{title}{Controlling reaction dynamics in chemical model
  systems through external driving},
\newblock \bibinfo{journal}{Physica D} \bibinfo{volume}{427}
  (\bibinfo{year}{2021}) \bibinfo{pages}{133013}.
  \DOIprefix\doi{10.1016/j.physd.2021.133013}.

\end{thebibliography}
\end{document}